\documentclass[showpacs,preprintnumbers,amsmath,amssymb,prb]{revtex4}

\usepackage{graphicx}
\usepackage{bm}

\begin{document}

\title{Spin-wave interaction in two-dimensional ferromagnets with dipolar forces}

\author{A. V. Syromyatnikov}
 \email{syromyat@thd.pnpi.spb.ru}
\affiliation{Petersburg Nuclear Physics Institute, Gatchina, St.\ Petersburg 188300, Russia}

\date{\today}

\begin{abstract}

We discuss the spin-wave interaction in two-dimensional (2D) Heisenberg ferromagnet (FM) with dipolar forces at $T_C\gg T\ge0$ using $1/S$ expansion. A comprehensive analysis is carried out of the first $1/S$ corrections to the spin-wave spectrum. In particular, similar to 3D FM discussed in our previous paper A.V. Syromyatnikov, PRB {\bf 74}, 014435 (2006), we obtain that the spin-wave interaction leads to the {\it gap} in the spectrum $\epsilon_{\bf k}$ renormalizing greatly the bare gapless spectrum at small momenta $k$. Expressions for the spin-wave damping $\Gamma_{\bf k}$ are derived self-consistently and it is concluded that magnons are well-defined quasi-particles in both quantum and classical 2D FMs at small $T$. We observe thermal enhancement of both $\Gamma_{\bf k}$ and $\Gamma_{\bf k}/\epsilon_{\bf k}$ at small momenta. In particular, a peak appears in $\Gamma_{\bf k}$ and $\Gamma_{\bf k}/\epsilon_{\bf k}$ at small $k$ and at any given direction of $\bf k$. If $S\sim1$ the height of the peak in $\Gamma_{\bf k}/\epsilon_{\bf k}$ is not larger than a value proportional to $T/D\ll1$, where $D$ is the spin-wave stiffness. In the case of large spins $S\gg1$ the peak in $\Gamma_{\bf k}/\epsilon_{\bf k}$ cannot be greater than that of the classical 2D FM found at $k=0$ which height is small only {\it numerically}: $\Gamma_{\bf 0}/\epsilon_{\bf 0}\approx0.16$ for the simple square lattice. Frustrating next-nearest-neighbor exchange coupling increases $\Gamma_{\bf 0}/\epsilon_{\bf 0}$ in classical 2D FM only slightly. We find expressions for spin Green's functions and the magnetization. The latter differs from the well-known result by S.V. Maleev, Sov. Phys. JETP {\bf 43}, 1240 (1976). The effect of the exchange anisotropy is also discussed briefly. Higher order corrections to the spectrum are considered and it is concluded that they are small compared to the first corrections obtained. Our results contradict to findings of the previous works (Ar. Abanov, A. Kashuba, V.L. Pokrovsky, PRL {\bf 77}, 2554 (1996); PRB {\bf 56}, 3181 (1997)) in which a diffusion spin-wave mode was obtained at small momenta. It is shown that the origin of this discrepancy is that the spin-wave gap was ignored in the previous studies.

\end{abstract}


\maketitle

\section{Introduction}
\label{int}

Magnetic properties of thin films and (quasi-)two-dimensional magnetic materials are of great interest now. \cite{rev,rev-jen} This interest is stimulated by recent advances in film growth techniques and numerous technological applications of magnetic films, including uses in electronics and data storage. A realistic theoretical model of low-dimensional magnetic systems must include the exchange interaction, the dipolar interaction, and the magnetocrystalline anisotropy. \cite{rev} Despite its smallness the long-range dipolar interaction plays essential role in 2D magnets. In particular, it violates the Mermin-Wagner theorem \cite{mw} and leads to stabilization of the long-range magnetic order at finite temperature in 2D magnets. \cite{rev,mal,2daf} Some peculiar features have been observed recently at $T\ll T_C$ related with the dipolar interaction both in 2D and 3D Heisenberg ferromagnets (FMs).

In 3D FM the problem of infrared singularities arose. It was obtained in Ref.~\cite{top} that dipolar forces lead to strong long-wavelength fluctuations manifesting themselves in infrared divergence of the first perturbation corrections to the uniform longitudinal spin susceptibility: $\chi_\|(\omega\to0) \sim iT/\omega$. An infrared divergent contribution to the spin-wave stiffness was obtained in Ref.~\cite{rahman} as a result of analysis of the first $1/S$ corrections to the spin-wave spectrum. Thus, the problem arose of analysis of the whole perturbation series in order to find spin-wave spectrum and longitudinal spin susceptibility at small momenta. Appearance of the infrared singularities in these papers is related to the fact that the spectrum is gapless in 3D FM in the spin-wave approximation. \cite{sw} 

First perturbation corrections to the spin-wave spectrum were analyzed in classical 2D FM with dipolar forces in Refs.~\cite{aprl,aprb}. It was found that the imaginary part of these corrections exceeds the bare gapless spectrum at small enough momenta $k$. Then as a result of self-consistent calculations a diffusion spin-wave mode was obtained at very small $k$. It was argued in Refs.~\cite{aprl,aprb} that despite the analysis is carried out for classical 2D FM the diffusion mode should be observed also in quantum 2D FM. 

Meantime we find that quite an unusual property of ferromagnets with dipolar forces is ignored in the previous studies that is crucial for the reported peculiarities --- interaction between magnons leads to appearance of a {\it gap} in the spin-wave spectrum that renormalizes the bare gapless spectrum greatly at small momenta. Thus, we obtain such gap in 3D FM in the first order of $1/S$ in our recent paper \cite{i}. It was found to be proportional to $\omega_0\sqrt{S\omega_0/J}\sin\theta_{\bf k}$, where 
\begin{equation}
\label{o0}
\omega_0 = 4\pi \frac{(g\mu)^2}{v_0}
\end{equation}
is the characteristic dipolar energy, $v_0$ is the unit cell volume, $\theta_{\bf k}$ is the angle between momentum $\bf k$ and magnetization and $J$ is the exchange value. We show that this gap screens the infrared singularities obtained in Refs.~\cite{top,rahman} and first perturbation corrections to the observables found self-consistently are small. Naturally, one can expect existence of such gap in 2D FM too.

It should be noted that appearance of the gap in 2D and 3D FMs is quite expected because dipolar interaction due to its symmetry and long-range nature violates the Goldstone theorem. Besides, the spin-wave gap was observed in 2D antiferromagnet with dipolar interaction in the zeroth order of $1/S$ (i.e., in the spin-wave approximation). \cite{2daf} Existence of the gap in the spectrum of 3D FM was anticipated long times ago. It is well known that within the first order of $1/S$ dipolar and pseudodipolar forces lead to anisotropic corrections to the total energy of 3D FM. \cite{tes,vleck,kef_an} As a result, directions along edges of the cube are energetically favorable in a simple cubic lattice whereas the magnetization should be parallel to a body diagonal of the cube in the face-centered cubic lattice and the body-centered cubic lattice. \cite{keffer,tes,vleck,kef_an} It has been pointed out by Keffer \cite{keffer,kef_an} that the anisotropic terms in the total energy of a ferromagnet should be accompanied with an "energy shift" in the spin-wave spectrum. We confirmed this long-standing statement for 3D FM in our previous paper \cite{i} and demonstrated the relation between the gap and the anisotropy at $T=0$. A four-fold in-plane anisotropy in square 2D FM caused by quantum fluctuations was also observed before and the value of this anisotropy was investigated numerically in Ref.~\cite{anis_dan} for $T\sim SJ$.

The situation is slightly different in classical FMs with dipolar interaction. In particular, the ground state remains infinitely degenerate according to the rotation about the axis perpendicular to the plane in the classical 2D FM. But it is well known that in classical systems with degenerate ground states some of these states can be selected via order-by-disorder mechanism at finite temperature. \cite{od1,od2,anis_sona} Order-by-disorder effect was demonstrated in 2D systems of classical spins confined to lie within the plane and coupled via (i) short-range dipolar-like interaction (Ref.~\cite{anis_sona}) and (ii) long-range dipolar interaction (Refs.~\cite{anis_bell,anis_it}). Then, an in-plane anisotropy arises at finite temperature. Notably, it was found in Ref.~\cite{anis_it} that the thermal selection of the ground state is accompanied by the appearance of the gap proportional to $T$ in the spin-wave spectrum originating from the spin-wave interaction and leading to the finite value of the order parameter. Then, one expects also appearance of the spin-wave gap in classical 2D FM with dipolar forces.

In the present paper we carry out a comprehensive analysis of the first $1/S$ corrections to the spin-wave spectrum in 2D FM with dipolar interaction. Similar to 3D FM we obtain the spin-wave gap $\Delta$ in the spectrum $\epsilon_{\bf k}$ which appears to be proportional to $\omega_0\sqrt{S\omega_0/J}$ for $S\sim1$ and $S\omega_0\sqrt{(\omega_0/J)(T/S^2J)}$ for $S\gg1$ and $T_C\gg T\gg SJ$. This gap renormalizes greatly the bare gapless spectrum at momenta $k\alt(\Delta/(S\omega_0))^2$. The limiting case of classical spins is also discussed at $T\ll j$ and the gap was found to be proportional to $w\sqrt{Tw/j^2}$, where $j$ and $w$ are values of the exchange and the characteristic dipolar energy in the classical model, respectively. It is demonstrated below that the spin-wave gap that was ignored in Refs.~\cite{aprl,aprb} are much larger than the energy of the diffusion mode obtained in those papers. We derive the spin-wave damping $\Gamma_{\bf k}$ self-consistently and find that spin waves are well-defined quasi-particles in both quantum and classical 2D FMs at small $T$. 

Interestingly, we observe thermal enhancement of the damping at small momenta. In particular, in quantum 2D FM we obtain a peak in both $\Gamma_{\bf k}$ and $\Gamma_{\bf k}/\epsilon_{\bf k}$ at $k\ll \omega_0/J$, $T\gg S\omega_0$ and at any given $\bf k$ direction. If $S\sim1$ the height of the peak in $\Gamma_{\bf k}/\epsilon_{\bf k}$ cannot axceed a value proportional to $T/(SJ)\ll1$. In the case of large spins $S\gg1$ the peak in $\Gamma_{\bf k}/\epsilon_{\bf k}$ cannot be greater than that of the classical 2D FM found at $k=0$ which height is small only {\it numerically}: $\Gamma_{\bf 0}/\epsilon_{\bf 0}\approx0.16$ for the simple square lattice. Frustrating next-nearest-neighbor exchange coupling increases $\Gamma_{\bf 0}/\epsilon_{\bf 0}$ in classical 2D FM only slightly.

We derive expressions for spin Green's functions and the magnetization. The latter differs from the well-known result of Ref.~\cite{mal}. The effect of the exchange anisotropy is also discussed briefly. Higher order corrections to the spectrum are considered and it is concluded that they are small compared to the first corrections obtained. We derive the four-fold in-plain anisotropy in the total energy of quantum 2D FM that makes directions of the magnetization along edges of the square to be energetically favorable in the simple square lattice. We also demonstrate the relation between this anisotropy and the spin-wave gap at $T=0$ as it was done in our previous paper \cite{i} for 3D FM.

The rest of the present paper is organized as follows. The Hamiltonian transformation and the technique are discussed in Sec.~\ref{hamtr}. First $1/S$ corrections to the real and imaginary parts of the spin-wave spectrum are considered in Secs.~\ref{specorr} and \ref{damping}, respectively. In Sec.~\ref{ls} we study the case of large spin values and consider the limit of classical spins. In Sec.~\ref{disc} we i) demonstrate the relation between the anisotropic term in the total energy appearing due to the dipolar interaction and the spin-wave gap at $T=0$, (ii) discuss the further order $1/S$ corrections to the spin-wave spectrum, (iii) calculate the magnetization taking into account the spin-wave spectrum renormalization, (iv) derive the spin Green's functions in the first order of $1/S$, and (v) discuss briefly the effect of the easy-plane anisotropy and consider the spectrum renormalization obtained in Refs.~\cite{aprl,aprb}. Sec.~\ref{con} contains a detailed summary and our conclusion. Three appendixes are included with some details of calculations.

\section{Hamiltonian transformation and technique}
\label{hamtr}

\begin{figure}
\centering
\includegraphics[scale=0.8]{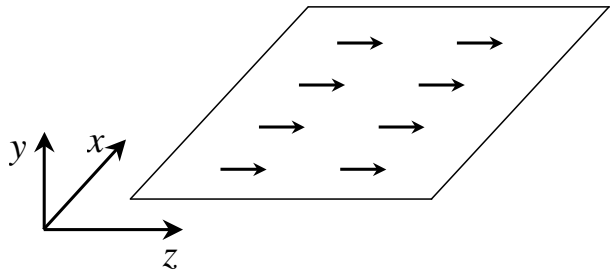}
\caption{Two-dimensional ferromagnet discussed in the present paper.
\label{plane}} 
\end{figure}

The Hamiltonian of a ferromagnet with dipolar interaction has the form
\begin{eqnarray}
\label{ham0}
{\cal H} &=& -\frac12 \sum_{l\ne m} \left(J_{lm}\delta_{\rho\beta} + Q_{lm}^{\rho\beta}\right) S_l^\rho  S_m^\beta,\\
\label{q}
Q_{lm}^{\rho\beta} &=& (g\mu)^2\frac{3R_{lm}^\rho R_{lm}^\beta - \delta_{\rho\beta}R_{lm}^2}{R_{lm}^5}.
\end{eqnarray}
Taking the Fourier transformation we have from Eq.~(\ref{ham0})
\begin{equation}
\label{ham}
{\cal H} = -\frac12 \sum_{\bf k}\left(J_{\bf k}\delta_{\rho\beta} + Q_{\bf k}^{\rho\beta}\right) S_{\bf k}^\rho  S_{-\bf k}^\beta,
\end{equation}
where $J_{\bf k} = \sum_l J_{lm}\exp(i{\bf k R}_{lm})$ and $Q_{\bf k}^{\rho\beta} = \sum_l Q_{lm}^{\rho\beta}\exp(i{\bf k R}_{lm})$. We direct $y$-axis perpendicular to the lattice as is shown in Fig.~\ref{plane}. Dipolar tensor $Q_{\bf k}^{\rho\beta}$ possesses the well-known properties \cite{mal,rev} at $k\ll1$, which are independent of the lattice type and the orientation of $x$ and $z$ axes relative to the lattice,
\begin{eqnarray}
\label{qsmall}
Q_{\bf k}^{\rho\beta} &=& \omega_0\left( \frac\alpha3 \delta_{\rho\beta} - \frac k2 \frac{k_\rho k_\beta}{k^2} \right), \mbox{ where } \rho,\beta=x,z,\\
\label{qy}
Q_{\bf k}^{y\beta} &=& \omega_0 \left( -\frac23 \alpha + \frac k2 \right)\delta_{y\beta},
\mbox{ where } \beta=x,y,z,\\
\alpha &=& \frac{3v_0}{8\pi}\sum_i \frac{1}{R_i^3},
\end{eqnarray}
where we set the lattice spacing to be equal to unity, $\omega_0$ is the characteristic dipolar energy given by Eq.~(\ref{o0}) and $\alpha$ is a constant that is equal approximately to $1.078$ for the simple square lattice. It is seen from Eqs.~(\ref{ham0}), (\ref{qsmall}) and (\ref{qy}) that dipolar forces lead to easy-plane anisotropy in the energy of the classical 2D FM with $y$ to be a hard axis. \cite{mal} We show in the next section that quantum and thermal fluctuations lead also to an in-plain anisotropy. We direct $z$-axis along the uniform magnetization as is shown in Fig.~\ref{plane}.

After Dyson-Maleev transformation
\begin{equation}
\label{dm}
S^x_{\bf k} = \sqrt{\frac S2} \left( a_{\bf k} + a^\dagger_{-\bf k} - \frac{(a^\dagger a^2)_{\bf k}}{2S} \right), \qquad 
S^y_{\bf k} = -i\sqrt{\frac S2} \left( a_{\bf k} - a^\dagger_{-\bf k} - \frac{(a^\dagger a^2)_{\bf k}}{2S} \right), \qquad  
S^z_{\bf k} = S - (a^\dagger a)_{\bf k}
\end{equation}
Hamiltonian (\ref{ham}) has the form ${\cal H} = E_0 + \sum_{i=1}^6 {\cal H}_i$, where $E_0$ is the ground state energy and ${\cal H}_i$ denote terms containing products of $i$ operators $a$ and $a^\dagger$. One should take into account terms up to ${\cal H}_4$ to calculate corrections of the first order in $1/S$. ${\cal H}_1=0$ because it contains only $Q_{\bf 0}^{\rho\beta}$ with $\rho\ne\beta$. For the remaining necessary terms one has
\begin{eqnarray}
\label{h2}
{\cal H}_2 &=& \sum_{\bf k} \left[E_{\bf k} a^\dagger_{\bf k}a_{\bf k} + \frac{B_{\bf k}}{2} \left(a_{\bf k}a_{-\bf k} + 
a^\dagger_{\bf k}a^\dagger_{-\bf k}\right)\right],\\
\label{h3}
{\cal H}_3 &=& \sqrt{\frac {S}{2 \mathfrak N}} \sum_{{\bf k}_1 + {\bf k}_2 + {\bf k}_3 = {\bf 0}} 
Q_2^{xz} a^\dagger_{-1}\left(a^\dagger_{-2} + a_2\right)a_3,\\
{\cal H}_4 &=& \frac{1}{4 \mathfrak N}\sum_{{\bf k}_1 + {\bf k}_2 + {\bf k}_3 + {\bf k}_4 = {\bf 0}} 
\left[
2\left(J_1-J_{1+3}\right)a^\dagger_{-1}a^\dagger_{-2}a_3a_4 \right.\nonumber\\
\label{h4}
&&{}\left.
+ a^\dagger_{-1} 
\left(
a_2 (Q_2^{xx} - Q_2^{yy}) + a^\dagger_{-2}\left( Q_2^{xx} + Q_2^{yy} - 2Q_{2+3}^{zz}\right) 
\right) 
a_3a_4 
\right],
\end{eqnarray}
where we drop index $\bf k$ in Eqs.~(\ref{h3}) and (\ref{h4}), $\mathfrak N$ is the number of spins in the lattice and
\begin{eqnarray}
\label{e}
E_{\bf k} &=& S(J_{\bf 0} - J_{\bf k}) - \frac S2\left( Q_{\bf k}^{xx} + Q_{\bf k}^{yy} - \frac{2\omega_0\alpha}{3} \right) \nonumber\\
& \stackrel{k\ll1}{\approx} & Dk^2 + \frac{S\omega_0\alpha}{2} - \frac{S\omega_0}{4}k\cos^2\phi_{\bf k},
\\
\label{b}
B_{\bf k} &=& \frac S2 \left( Q_{\bf k}^{yy} - Q_{\bf k}^{xx} \right)
\stackrel{k\ll1}{\approx} -\frac{S\omega_0\alpha}{2} + \frac{S\omega_0}{4}k (1 + \sin^2\phi_{\bf k}),
\end{eqnarray}
where $D$ is the spin-wave stiffness, $\phi_{\bf k}$ is the angle between $\bf k$ and the magnetization and the expressions after $\stackrel{k\ll1}{\approx}$ are approximate values of the corresponding quantities at $k\ll1$. One has for the coupling between only nearest neighbor spins on the simple square lattice $D=SJ$. In the spin-wave approximation we find for the magnon spectrum
\begin{equation}
\label{spec1}
\epsilon_{\bf k} = \sqrt{E_{\bf k}^2 - B_{\bf k}^2} \stackrel{k\ll1}{\approx} \sqrt{\left(Dk^2 + S\omega_0\alpha \right)\left(Dk^2 + \frac{S\omega_0}{2}k\sin^2\phi_{\bf k}\right)}
\end{equation}
in accordance with the well-known result. \cite{mal}

To perform the calculations it is convenient to introduce the following retarded Green's functions: $G(\omega,{\bf k}) = \langle a_{\bf k}, a^\dagger_{\bf k} \rangle_\omega$, $F(\omega,{\bf k}) = \langle a_{\bf k}, a_{-\bf k} \rangle_\omega$, ${\overline G}(\omega,{\bf k}) = \langle a^\dagger_{-\bf k}, a_{-\bf k} \rangle_\omega = G^*(-\omega,-{\bf k})$ and $F^\dagger (\omega,{\bf k}) = \langle a^\dagger_{-\bf k}, a^\dagger_{\bf k} \rangle_\omega = F^*(-\omega,-{\bf k})$. We have two sets of Dyson equations for them. One of these sets has the form
\begin{equation}
\label{eqfunc}
\begin{array}{l}
G(\omega,{\bf k}) = G^{(0)}(\omega,{\bf k}) + G^{(0)}(\omega,{\bf k}){\overline \Sigma}(\omega,{\bf k})G(\omega,{\bf k}) + G^{(0)}(\omega,{\bf k}) [B_{\bf k} + \Pi(\omega,{\bf k})] F^\dagger(\omega,{\bf k}),\\
F^\dagger(\omega,{\bf k}) = {\overline G}^{(0)}(\omega,{\bf k}) \Sigma(\omega,{\bf k})F^\dagger(\omega,{\bf k}) + {\overline G}^{(0)}(\omega,{\bf k}) [B_{\bf k} + \Pi^\dagger(\omega,{\bf k}) ]G(\omega,{\bf k}),
\end{array}
\end{equation}
where $G^{(0)}(\omega,{\bf k}) = (\omega - E_{\bf k}+i\delta)^{-1}$ is the bare Green's function and $\Sigma$, $\overline \Sigma$, $\Pi$ and $\Pi^\dagger$ are the self-energy parts. Solving Eqs.~(\ref{eqfunc}) one obtains
\begin{eqnarray}
G(\omega,{\bf k}) &=& \frac{\omega + E_{\bf k} + \Sigma(\omega,{\bf k})}{{\cal D}(\omega,{\bf k})},\nonumber\\
F(\omega,{\bf k}) &=& -\frac{B_{\bf k} + \Pi(\omega,{\bf k})}{{\cal D}(\omega,{\bf k})},\nonumber\\
\label{gf}
{\overline G}(\omega,{\bf k}) &=& \frac{-\omega + E_{\bf k} + {\overline \Sigma}(\omega,{\bf k})}{{\cal D}(\omega,{\bf k})},\\
F^\dagger(\omega,{\bf k}) &=& -\frac{B_{\bf k} + \Pi^\dagger(\omega,{\bf k})}{{\cal D}(\omega,{\bf k})},\nonumber
\end{eqnarray}
where
\begin{eqnarray}
\label{d}
{\cal D}(\omega,{\bf k}) &=& (\omega+i\delta)^2 - \epsilon_{\bf k}^2 - \Omega(\omega,{\bf k}),\\
\label{o}
\Omega(\omega,{\bf k}) &=& E_{\bf k}(\Sigma + \overline{\Sigma}) - B_{\bf k}(\Pi + \Pi^\dagger) - (\omega + i\delta)(\Sigma - \overline{\Sigma}) - \Pi\Pi^\dagger + \Sigma \overline{\Sigma},
\end{eqnarray}
and $\epsilon_{\bf k}$ is given by Eq.~(\ref{spec1}). Quantity $\Omega(\omega,{\bf k})$ given by Eq.~(\ref{o}) describes renormalization of the spin-wave spectrum square. We calculate the real part of $\Omega(\omega,{\bf k})$ in the next section and analyze its imaginary part in Sec.~\ref{damping}. The last two terms in Eq.~(\ref{o}) give corrections of at least second order in $1/S$ and are not considered in Secs.~\ref{specorr}--\ref{ls}. We imply that $S\sim1$ in the next two sections and discuss large $S$ in Sec.~\ref{ls}.

\section{Renormalization of the real part of the spin-wave spectrum}
\label{specorr}

Corrections to the spin-wave spectrum to be obtained are proportional to sums over momenta in which summands depend on the components of the dipolar tensor $Q_{\bf k}^{\rho\beta}$. In some of these sums summation over small momenta is important and one can use expressions (\ref{qsmall}) and (\ref{qy}) for $Q_{\bf k}^{\rho\beta}$. Meantime there are sums in which summation over large momenta is essential and which, consequently, depend on the direction of the quantized axis within the plane and the lattice type. Thus, one should bear in mind what is the direction of magnetization in the ground state. 

Therefore the well-known fact should be taken into account that dipolar and pseudodipolar interactions lead to the dependence of the energy of a ferromagnet on the direction of quantized axis. \cite{tes,kef_an} We show now that, being established first for 3D FM, this finding remains also valid for 2D FM: apart from the easy-plain anisotropy found in Ref.~\cite{mal} and discussed above there is also an in-plain anisotropy. As in 3D FM, the first $1/S$-correction to the classical energy $E_0$ having the form
\begin{equation}
\label{de}
\Delta E = \langle {\cal H}_2\rangle = \sum_{\bf k} \frac{\epsilon_{\bf k} - E_{\bf k}}{2} 
\approx - \sum_{\bf k} \frac{B_{\bf k}^2}{2 (\epsilon_{\bf k} + E_{\bf k})}
\end{equation}
gives rise to such in-plain anisotropy, where the isotropic term is omitted in the right part of Eq.~(\ref{de}). After direct calculations one obtains for a square lattice
\begin{eqnarray}
\label{an}
\frac{\Delta E}{\mathfrak N} &=& C \frac{(S\omega_0)^2}{2D} \gamma_x^2\gamma_z^2,\\
\label{c}
C &=& \frac{D}{\omega_0^2 \mathfrak N} \sum_{\bf q}\frac{\left( Q_{\bf q}^{xx} - Q_{\bf q}^{zz} \right)^2 - 4 \left( Q_{\bf q}^{xz} \right)^2 }{8\epsilon_{\bf q}},
\end{eqnarray}
where $\gamma_i$ are direction cosines of the magnetization and components of the dipolar tensor in Eq.~(\ref{c}) are taken relative to square axes. The constant $C$ should be calculated numerically because summation over large momenta is important in Eq.~(\ref{c}) and one cannot use Eqs.~(\ref{qsmall}) and (\ref{qy}) for the dipolar tensor components. This calculation can be carried out using the dipolar sums computation technique (see, e.g., Ref.~\cite{sums} and references therein) with the result $C\approx0.0082$ for exchange coupling between only nearest neighbor spins on the simple square lattice. Because $C>0$ an edge of the square is the easy direction. Notice also that the value of the dipolar anisotropy in a cubic 3D FM is proportional to a constant that also has the form (\ref{c}), where, naturally, summation is taken over a 3D lattice. \cite{i} The in-plane anisotropy caused by dipolar interaction was investigated numerically at $T\sim D$ in Ref.~\cite{anis_dan}.

\begin{figure}
\centering
\includegraphics[scale=0.8]{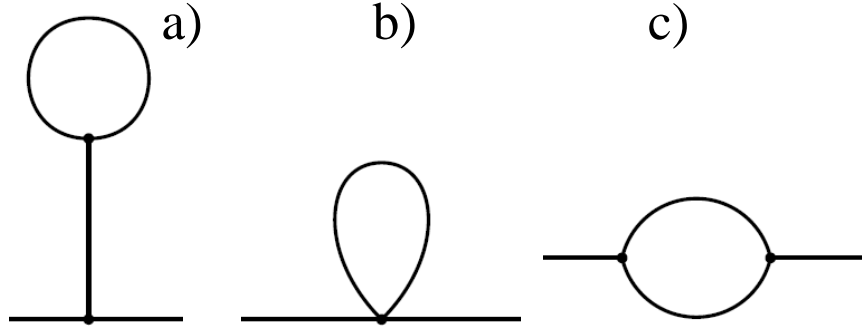}
\caption{Diagrams of the first order in $1/S$ for self-energy parts discussed in this paper. Diagrams (a) and (c) stem from three-magnon terms (\ref{h3}) in the Hamiltonian whereas (b) comes from four-magnon terms (\ref{h4}).
\label{diag}} 
\end{figure}

We study now separately bubble diagrams shown in Fig.~\ref{diag}(a) and (b), and the loop diagram presented in Fig.~\ref{diag}(c). 

\subsection{Bubble diagrams}
\label{bd}

Let us start with the diagram shown in Fig.~\ref{diag}(a). It appears from three-magnons terms (\ref{h3}) and gives zero. To demonstrate this we make all possible couplings of two operators $a$ and $a^\dagger$ in Eq.~(\ref{h3}): 
\begin{equation}
\label{h3mod}
\left(a_{\bf 0}^\dagger + a_{\bf 0}\right)
\sqrt{\frac {S}{2 \mathfrak N}}
\sum_{\bf q} \frac{(E_{\bf q} - B_{\bf q})(1+2N_{\bf q}) - \epsilon_{\bf q}}{2\epsilon_{\bf q}}Q_{\bf q}^{xz},
\end{equation}
where $N_{\bf q}=(e^{\epsilon_{\bf q}/T}-1)^{-1}$ is Plank's function. Expression (\ref{h3mod}) is equal to zero because $Q_{\bf q}^{xz} = -Q_{\bf q'}^{xz}$, $E_{\bf q} = E_{\bf q'}$ and $B_{\bf q} = B_{\bf q'}$, where ${\bf q}=(q_x,q_z)$ and ${\bf q'}=(-q_x,q_z)$.

The Hartree-Fock diagram presented in Fig.~\ref{diag}(b) comes from ${\cal H}_4$-terms given by Eq.~(\ref{h4}). After simple calculations we obtain for the contribution to $\Omega(\omega,{\bf k})$ from this diagram
\begin{eqnarray}
\label{o4}
\Omega^{(4)}(\omega,{\bf k}) &=& 
\frac{E_{\bf k}}{\mathfrak N} \sum_{\bf q}\left[\frac{E_{\bf q}(1+2N_{\bf q}) - \epsilon_{\bf q}}{\epsilon_{\bf q}} \left(
J_{\bf k} - J_0 + J_{\bf q} - J_{\bf k+q} 
+ \frac12 \left( Q_{\bf q}^{xx} + Q_{\bf q}^{yy} + Q_{\bf k}^{xx} + Q_{\bf k}^{yy} - 2Q_{\bf k+q}^{zz} 
- 2Q_0^{xx} \right)
\right)
\right.
\nonumber\\
&&\left.
{}- \frac{B_{\bf q}(1+2N_{\bf q})}{4\epsilon_{\bf q}} \left( Q_{\bf k}^{xx} - Q_{\bf k}^{yy} + 2 \left(Q_{\bf q}^{xx} - Q_{\bf q}^{yy}\right)\right)
\right]
\nonumber\\
&&{}- 
\frac{B_{\bf k}}{ \mathfrak N} \sum_{\bf q}
\left[
\frac{E_{\bf q}(1+2N_{\bf q}) - \epsilon_{\bf q}}{2\epsilon_{\bf q}} 
\left(
\frac12 \left( Q_{\bf q}^{xx} - Q_{\bf q}^{yy} \right) + Q_{\bf k}^{xx} - Q_{\bf k}^{yy} 
\right)
\right.
\nonumber\\
&&\left.
{}- \frac{B_{\bf q}(1+2N_{\bf q})}{2\epsilon_{\bf q}} \left( J_{\bf k} + J_{\bf q} - 2J_{\bf k+q} 
+ \frac12 \left( Q_{\bf q}^{xx} + Q_{\bf q}^{yy} + Q_{\bf k}^{xx} + Q_{\bf k}^{yy} - 4Q_{\bf k+q}^{zz} \right)
\right)
\right].
\end{eqnarray}
At zero temperature $N_{\bf q}=0$ in Eq.~(\ref{o4}). In this case the spectrum is renormalized by quantum fluctuations only. As the temperature increases, corrections from terms in Eq.~(\ref{o4}) containing $N_{\bf q}$ become larger. They exceed terms in Eq.~(\ref{o4}) not containing $N_{\bf q}$ above a certain temperature. We find below that this temperature is of the order of $S\omega_0$. Then, it is convenient to discuss separately regimes $T\ll S\omega_0$ and $T\gg S\omega_0$.

\subsubsection{$T\ll S\omega_0$}

Taking $N_{\bf q}=0$ in Eq.~(\ref{o4}), we obtain in the leading order of $\omega_0$
\begin{equation}
\label{o4t<}
\Omega^{(4)}(\omega,{\bf k}) =
Dk^2  \frac{S\omega_0 \alpha}{2\mathfrak N} 
\sum_{\bf q}\frac{Q_{\bf q}^{xx} - Q_{\bf q}^{yy}}{\epsilon_{\bf q}} 
+ \frac{S^2\omega_0\alpha}{8\mathfrak N} 
\sum_{\bf q}\frac{\left(Q_{\bf q}^{xx} - Q_{\bf q}^{yy}\right)^2}{\epsilon_{\bf q}}.
\end{equation}
Here the first term is of the order of $k^2\omega_0^2\ln(D/(S\omega_0))$. Then it gives positive negligibly small correction to the bare spectrum (\ref{spec1}). In contrast, the second term in Eq.~(\ref{o4t<}), being independent of $\bf k$, contributes to the spin-wave gap. It is much greater than the bare spectrum at $k\ll S\omega_0/D$.

\subsubsection{$T\gg S\omega_0$}

Terms in Eq.~(\ref{o4}) containing $N_{\bf q}$ come into play at such $T$ and we have in the leading order of $\omega_0$
\begin{eqnarray}
\label{o4t>}
\Omega^{(4)}(\omega,{\bf k}) &=& 
- \left(Dk^2\right)^2\frac2S W(T) 
- Dk^2\omega_0 \alpha [W(T)+V(T)]
+ \frac{S^2\omega_0\alpha}{8\mathfrak N} 
\sum_{\bf q}\frac{\left(Q_{\bf q}^{xx} - Q_{\bf q}^{yy}\right)^2}{\epsilon_{\bf q}},\\
\label{w}
W(T) &=& \frac{1}{\mathfrak N} \sum_{\bf q} \frac{J_{\bf 0} - J_{\bf q}}{J_{\bf 0}}  N_{\bf q} \approx w \left(\frac TD\right)^2,\\
\label{v}
V(T) &=& \frac{2}{\mathfrak N} \sum_{\bf q} N_{\bf q} \approx v \frac TD \ln\left(\frac{T}{S\omega_0}\right),
\end{eqnarray}
where $w = (16\pi)^{-1}\int_0^\infty dk k/(e^k-1) = \pi/96$ and $v = 1/(2\pi)$. The first two terms in Eq.~(\ref{o4t>}) do not change the structure of the bare spectrum resulting in the following renormalization of constants $D$ and $\omega_0$ in Eq.~(\ref{spec1}):
\begin{equation}
\label{ren}
D \mapsto D\left[1 - \frac 1S W(T)\right],\qquad
\omega_0 \mapsto \omega_0 \left[1 - \frac 1S V(T)\right].
\end{equation}
In contrast, the last term in Eq.~(\ref{o4t>}) changes the form of the spectrum and contributes to the gap. Notice that the gap in Eq.~(\ref{o4t>}) has the same form as in Eq.~(\ref{o4t<}). Thermal corrections to the gap are small being of the order of $\omega_0^3T\ln(T/(S\omega_0))/D^2$.
\footnote{
Thermal corrections to the gap proportional to $\omega_0^3T\ln(T/(S\omega_0))/D^2$ from the Hartree-Fock diagram and from the loop diagram discussed in Sec.~\ref{ld} cancel each other. As a result thermal correction to the square of the gap for $S\sim1$ is much smaller, $\omega_0^3T^{3/2}/D^{5/2}$, (see discussion in Sec.~\ref{re}). To find it one should take into account ${\cal O}(k^2)$ corrections in the expansion of the dipolar tensor $Q_{\bf k}^{\rho\beta}$ which depend on the lattice type.
}

Comparing first two terms in Eq.~(\ref{o4t>}) with the first term in Eq.~(\ref{o4t<}) one infers that thermal $k$-dependent corrections become much larger than quantum ones at $T\gg S\omega_0$. 

\subsection{Loop diagram}
\label{ld}

We turn now to the loop diagram shown in Fig.~\ref{diag}(c). It originates from ${\cal H}_3$-terms (\ref{h3}) in the Hamiltonian. As a result of simple but tedious calculations some details of which are presented in Appendix~\ref{loop} we have for the contribution to the real part of $\Omega(\omega,{\bf k})$ from this diagram at $k\ll1$
\begin{eqnarray}
\label{o3t}
{\rm Re}\Omega^{(3)}(\omega,{\bf k}) &=& 
-
\frac{S^2\omega_0\alpha}{2\mathfrak N} \sum_{\bf q} 
\frac{ \left( Q_{\bf q}^{xz} \right)^2 }{\epsilon_{\bf q}},
\end{eqnarray}
where we set $k=0$ under the sum because summation over large $\bf q$ is essential. Notice that we discard in Eq.~(\ref{o3t}) all the terms that are much smaller than $\Omega^{(4)}(\omega,{\bf k})$ given by Eqs.~(\ref{o4t<}) and (\ref{o4t>}). Temperature corrections to ${\rm Re}\Omega^{(3)}(\omega,{\bf k})$ are negligible. 

\subsection{Resulting expressions}
\label{re}

One can derive now the resulting expression for ${\rm Re}\Omega(\omega,{\bf k})$ using Eqs.~(\ref{o4t<}), (\ref{o4t>}) and (\ref{o3t}).

\subsubsection{$T \ll S\omega_0$} 

\begin{figure}
\centering
\includegraphics{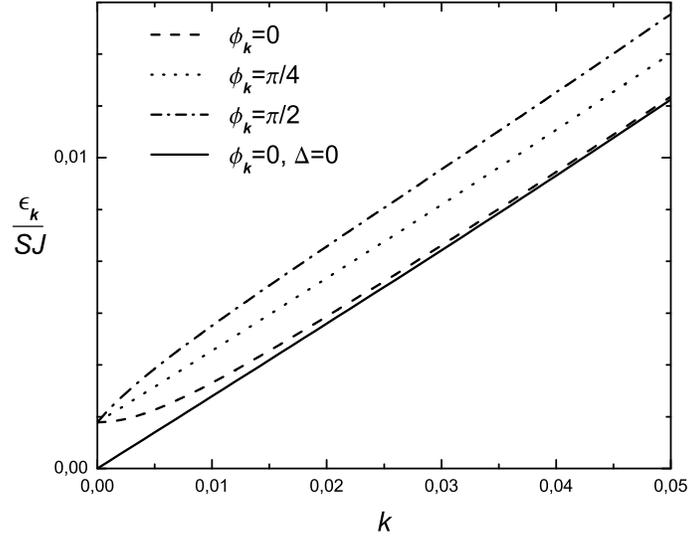}
\caption{Renormalized spin-wave spectrum at $\phi_{\bf k}=0$, $\phi_{\bf k}=\pi/4$ and $\phi_{\bf k}=\pi/2$ for a 2D FM at $T\ll S\omega_0$ with $\omega_0=0.05J$ and $S=1/2$. The bare spectrum at $\phi_{\bf k}=0$ (solid line) is also presented for comparison.
\label{spec2d}} 
\end{figure}

We obtain from Eqs.~(\ref{o4t<}) and (\ref{o3t})
\begin{eqnarray}
\label{ot1}
{\rm Re}\Omega(\omega,{\bf k}) &=& 
Dk^2  \frac{S\omega_0 \alpha}{2\mathfrak N} 
\sum_{\bf q}\frac{Q_{\bf q}^{xx} - Q_{\bf q}^{yy}}{\epsilon_{\bf q}} + \Delta^2,\\
\label{delta}
\Delta &=& \sqrt{\alpha CS^2\frac{\omega_0^3}{D}},
\end{eqnarray}
where $C$ is given by Eq.~(\ref{c}). As it is mentioned above, $C\approx0.0082$ for the simple square lattice with the exchange coupling between nearest neighbor spins only. The first term in Eq.~(\ref{ot1}) originates from $\Omega^{(4)}(\omega,{\bf k})$ and to the second term contribute both $\Omega^{(3)}(\omega,{\bf k})$ and $\Omega^{(4)}(\omega,{\bf k})$. The spin-wave gap $\Delta$ given by Eq.~(\ref{delta}) is proportional to $\omega_0^{3/2}$. To illustrate this result we plot in Fig.~\ref{spec2d} renormalized and the bare spin-wave spectrum for $\phi_{\bf k}=0$, $\phi_{\bf k}=\pi/4$ and $\phi_{\bf k}=\pi/2$ assuming that there is exchange coupling between nearest neighbor spins only, $\omega_0=0.05J$ and $S=1/2$.

\subsubsection{$T \gg S\omega_0$}

One has from Eqs.~(\ref{o4t>}) and (\ref{o3t})
\begin{eqnarray}
\label{ot2}
{\rm Re}\Omega(\omega,{\bf k}) &=& 
- \left(Dk^2\right)^2\frac2S W(T) 
- Dk^2\omega_0 \alpha [W(T)+V(T)] 
+ \Delta^2,
\end{eqnarray}
where $W(T)$, $V(T)$ and $\Delta$ are given by Eqs.~(\ref{w}), (\ref{v}) and (\ref{delta}), respectively. Notice that the spin-wave gap $\Delta$ has the same form as at $T\ll S\omega_0$, thermal corrections to its square are negligible being of the order of $\omega_0^3T^{3/2}/D^{5/2}$. 

We infer comparing Eq.~(\ref{spec1}) with Eqs.~(\ref{ot1}) and (\ref{ot2}) that renormalization of the bare spectrum is small at $T\ll D$ and $k\gg S\omega_0/D$. In contrast, the spectrum renormalization is significant at smaller $k$ due to the gap that is much larger than all $k$-dependent terms at $k\alt(\Delta/(S\omega_0))^2$. We use renormalized spectrum below for self-consistent calculation of the spin-wave damping and estimation of higher order $1/S$ corrections.

\section{Spin-wave damping}
\label{damping}

We discuss in this section the imaginary part of $\Omega(\omega,{\bf k})$ to which only the loop diagram contributes shown in Fig.~\ref{diag}(c). Corresponding calculations are rather cumbersome and we discuss only results here. One refers to Appendix~\ref{loop} for some details of the calculations. We discuss 2D FM on the simple square lattice in this section. Three regions should be considered: $k\gg\sqrt{S\omega_0/D}$, $S\omega_0/D\ll k\ll\sqrt{S\omega_0/D}$ and $k\alt(\Delta/(S\omega_0))^2$ at which the real part of the spectrum has the from $\epsilon_{\bf k}\approx Dk^2$, $\epsilon_{\bf k}\approx k \sqrt{\alpha SD\omega_0}$ and $\epsilon_{\bf k}\approx \Delta$, respectively. ${\rm Im}\Omega(\omega,{\bf k})$ is an odd function of $\omega$ and we calculate it for $\omega=\epsilon_{\bf k}$ only. The spin-wave damping $\Gamma_{\bf k}$ at momentum $\bf k$ is found below using the relation 
\begin{equation}
\label{dam}
\Gamma_{\bf k} = -\frac{{\rm Im}\Omega(\omega=\epsilon_{\bf k},{\bf k})}{2\epsilon_{\bf k}}.
\end{equation}

\subsection{$k\gg\sqrt{S\omega_0/D}$}

One obtains for $k\gg\sqrt{S\omega_0/D}$
\begin{equation}
\label{damq>}
\Gamma_{\bf k} = 
\epsilon_{\bf k}
\frac{S\omega_0^2}{D^2}\frac{1}{3\pi 2^8}
\left(
\left[
1+
24 \frac{T}{S\omega_0k}f(\phi_{\bf k})
\right] \sin^22\phi_{\bf k}
+
\frac{3\pi-8}{4}
+
6(\pi - 2) \frac{T}{Dk^2}
\right),
\end{equation}
where
\begin{equation}
\label{f}
f(\phi_{\bf k}) = \int_0^\infty \frac{dq}{q^2 + q \cos^2\phi_{\bf k} + 4D\Delta^2/(\alpha S^3\omega_0^3)}
\end{equation}
and $\Delta$ is given by Eq.~(\ref{delta}). It is seen that the damping is anisotropic in this regime: it is smaller along edges of the square and it reaches maxima along diagonals of the square. 

\subsection{$S\omega_0/D \ll k\ll\sqrt{S\omega_0/D}$}

\begin{figure}
\centering
\includegraphics{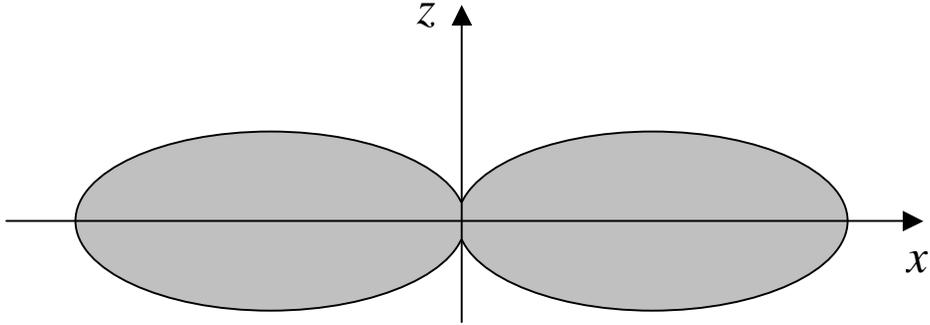}
\caption{The region in $\bf k$-plane is shown in which the spin-wave damping is zero at $T=0$. This region is defined by Eq.~(\ref{reg}). The linear dimension of the region is of the order of $(S\omega_0/D)^{2/3}$. The length of the region along $z$ axis at $k_x=0$ is of the order of $(S\omega_0/D)^{3/4}$. As is shown in Fig.~\ref{plane}, $z$ axis is directed along magnetization. Great thermal enhancement of the damping inside this region at $T\gg S\omega_0$ is illustrated by Figs.~\ref{fdt} and \ref{fdinf}.
\label{damp}} 
\end{figure}

One has for $S\omega_0/D \ll k\ll\sqrt{S\omega_0/D}$
\begin{eqnarray}
\label{damq<}
\Gamma_{\bf k} &=&
\epsilon_{\bf k}
\frac{S^2 \omega_0^3}{D^3 k^2}
\frac{\alpha}{2^6 \sqrt3 \pi} 
\left(
\left[
{\cal A}_1({\bf k})
+
\frac{T}{\epsilon_{\bf k}}({\cal A}_2({\bf k}) + {\cal A}_3({\bf k}))
\right]
\sin^22\phi_{\bf k}\right.\nonumber\\
&&{}+
\left.
\frac{1}{20\alpha^3}\frac{D^3}{S^3\omega_0^3}k^6 
\left[
{\cal B}_1({\bf k}) - 3{\cal B}_2({\bf k}) + \frac94{\cal B}_3({\bf k}) 
\right]
+
\frac{T}{S\omega_0\alpha}{\cal B}_4
\right),
\end{eqnarray}
where
\begin{eqnarray}
{\cal A}_1({\bf k}) &=& \int_{-\sqrt{1-\varepsilon_1}}^{\sqrt{1-\varepsilon_1}} \frac{dq}{\sqrt{{\cal L}(q,{\bf k})}},\\
{\cal A}_2({\bf k}) &=& 4\int_{-\sqrt{1-\varepsilon_1}}^{\sqrt{1-\varepsilon_1}}
\frac{dq}{\left(1-q^2\right) \sqrt{{\cal L}(q,{\bf k})}},\\
{\cal A}_3({\bf k}) &=& 2\int_{\sqrt{1+\varepsilon_2}}^\infty dq
\frac{q+1}{q-1}
\frac{1}{\sqrt{{\cal L}(q,{\bf k})}},\\
{\cal B}_i({\bf k}) &=& \frac{15}{4} \int_{-\sqrt{1-\varepsilon_1}}^{\sqrt{1-\varepsilon_1}} dq
\frac{q^2}{\left(1-q^2\right)^{2i-2}} 
{\cal L}(q,{\bf k})^{(2i-1)/2}, \quad i=1,2,3,\\
\label{b4}
{\cal B}_4 &=& 4\sqrt3 \int_0^\infty dq\frac{(1+q^2)^{3/2} \sqrt{3+4q^2}}{(1+2q^2)^4} \approx 5.31,\\
\label{l}
{\cal L}(q,{\bf k}) &=& \left(1-q^2\right)^2 - \beta \left|1-q^2\right| - \xi \left(q^2+3\right),\\
\label{beta}
\beta &=& \frac{2\alpha}{3} \frac{S^2\omega_0^2}{D^2k^3} \sin^2\phi_{\bf k},\\
\label{gamma}
\xi &=& \frac43 \frac{\Delta^2}{D^2k^4},\\
\label{eps}
\varepsilon_1 &=& \frac12 \left(\beta-\xi + \sqrt{(\beta-\xi)^2 + 16\xi}\right),\\
\varepsilon_2 &=& \frac12 \left(\beta+\xi + \sqrt{(\beta+\xi)^2 + 16\xi}\right),
\end{eqnarray}
and ${\cal A}_{1,2}({\bf k})$ and ${\cal B}_{1,2,3}({\bf k})$ should be taken equal to zero at $\bf k$ such that $\varepsilon_1\ge1$, i.e., when the following inequality satisfies:
\begin{equation}
\label{reg}
1-\beta-3\xi\le0. 
\end{equation}
Notice that ${\cal L}(q,{\bf k}) = 0$ at $q=\pm\sqrt{1-\varepsilon_1},\sqrt{1+\varepsilon_2}$ and it is positive inside the intervals $(-\sqrt{1-\varepsilon_1},\sqrt{1-\varepsilon_1})$ and $(\sqrt{1+\varepsilon_2},\infty)$. When $\beta,\xi\ll1$ one has ${\cal A}_1({\bf k}) \approx -\ln(\beta+\sqrt\xi)$, ${\cal A}_{2,3}({\bf k}) \sim 1/(\beta+\sqrt\xi)$ and ${\cal B}_{1,2,3}({\bf k})\approx1$. If $\beta,\xi\agt 1$ we have ${\cal A}_3({\bf k}) \sim 1/\sqrt{\beta+\xi}$. Terms in Eq.~(\ref{damq<}) containing ${\cal B}_i({\bf k})$ and ${\cal B}_4$ play only at $|\sin2\phi_{\bf k}|\ll1$. In particular, we obtain from Eq.~(\ref{damq<}) that the spin-wave damping is zero at $T=0$ and such $\bf k$ that $\varepsilon_1\ge1$, i.e., when inequality (\ref{reg}) holds. The region determined by Eq.~(\ref{reg}) is sketched in Fig.~\ref{damp}.

We conclude from Eqs.~(\ref{damq>}) and (\ref{damq<}) that the damping is small in the corresponding intervals provided that $\omega_0,T\ll D$.

\subsection{$k \ll S\omega_0/D$}
\label{ksm}

As it is found above, the real part of the spectrum at $k\alt(\Delta/(S\omega_0))^2$ is renormalized greatly being equal approximately to $\Delta$. We find self-consistently for $S\omega_0/D \gg k \agt \Delta/\sqrt{TD}$
\begin{equation}
\label{damq<<}
\Gamma_{\bf k} = \frac{\alpha}{2^8}\frac{T\omega_0^3S^2}{D^2\epsilon_{\bf k}}.
\end{equation}
Notice that this regime is realized at large enough temperature, $T\gg \omega_0$. At smaller $k$, $k\ll\Delta/\sqrt{TD}\ll S\omega_0/D$, or at small temperature, $T\ll \omega_0$, the spin-wave damping is exponentially small:
\begin{equation}
\label{damq<<<}
\Gamma_{\bf k} \propto \exp\left(-\frac{\Delta^2}{4TDk^2}\right). 
\end{equation}
It should be noted from Eqs.~(\ref{damq<}) and (\ref{damq<<}) that when $T\gg S\omega_0$ the damping $\Gamma_{\bf k}$ increases upon decreasing $k$ for $k \alt (S\omega_0/D)^{2/3}$. On the other hand, as we have just obtained, the damping is exponentially small at very small momenta $k\ll\Delta/\sqrt{TD}$. Hence, one should observe a peak in $\Gamma_{\bf k}$ at $k\sim \Delta/\sqrt{TD}$ and at any $\bf k$ direction which height can be estimated from Eq.~(\ref{damq<<}). In particular, if the temperature is as large as the interval $(\Delta/(S\omega_0))^2\gg k\agt\Delta/\sqrt{TD}$ is finite, i.e., if $T\gg S^2\omega_0/C$, we have $\epsilon_{\bf k}\approx\Delta$ at $k\sim \Delta/\sqrt{TD}$ in Eq.~(\ref{damq<<}) and one obtains for the peak height from Eq.~(\ref{damq<<}) using Eq.~(\ref{delta})
\begin{equation}
\label{ratio}
\frac{\Gamma_{\bf k}}{\epsilon_{\bf k}} 
=
\frac{1}{2^8 C} \frac{T}{D}.
\end{equation}
In general case the peak height cannot be larger than the value given by Eq.~(\ref{ratio}) because $\epsilon_{\bf k}\ge\Delta$ at $k\ll S\omega_0/D$. Notice also that Eq.~(\ref{ratio}) is valid for arbitrary $\phi_{\bf k}$. 

It is seen from Eq.~(\ref{ratio}) that the spin-wave damping is much smaller than the real part of the spectrum if $T\ll D,T_C$, where
\begin{equation}
\label{tc}
T_C = \frac{4\pi DS}{\ln \left(4\pi S (D/[S\omega_0])^{3/2}\right)}
\end{equation}
is the value of the Curie temperature for $S\sim1$ obtained using the spin-wave theory (see Sec.~\ref{disc}). 
\footnote{
Although $T_C$ given by Eq.~(\ref{tc}) can differ several times from the real value of the Curie temperature, \cite{2dtc} this precision is sufficient for our estimations. 
} 
On the other hand the temperature can be greater than $D$ for large spin values much greater than unity, $S\agt\ln \left(4\pi S (D/[S\omega_0])^{3/2}\right)$,
\footnote{
Notice that even at large enough dipolar characteristic energy $\omega_0$ (remaining small compared to the exchange value), say at $\omega_0\approx0.1J$, we have $S\agt\ln \left(4\pi S (D/[S\omega_0])^{3/2}\right)$ only for pretty large spins $S\agt8$.
} 
so that $D< T\ll T_C^{(S\gg1)}$, where 
\begin{equation}
\label{tci}
T_C^{(S\gg1)} = \frac{8\pi DS}{3\ln(D/[S\omega_0])}
\end{equation}
is the Curie temperature in the spin-wave approximation for $S\gg\ln \left(4\pi S (D/[S\omega_0])^{3/2}\right)$. Then one might conclude from Eq.~(\ref{ratio}) that the damping can be much larger than the real part of the spectrum. Meantime we show in the next section that the temperature correction to the spin-wave gap is large at such large $T$ and $S$. As a result the imaginary part of the spectrum is also much smaller than the real part at $S\gg1$ and $D<T\ll T_C^{(S\gg1)}$ and the peak height of the ratio $\Gamma_{\bf k}/\epsilon_{\bf k}$ in quantum 2D FM cannot be larger than that in the classical 2D FM which is equal approximately to $0.16$ for the simple square lattice with exchange coupling between nearest spins only.

\begin{figure}
\centering
\includegraphics{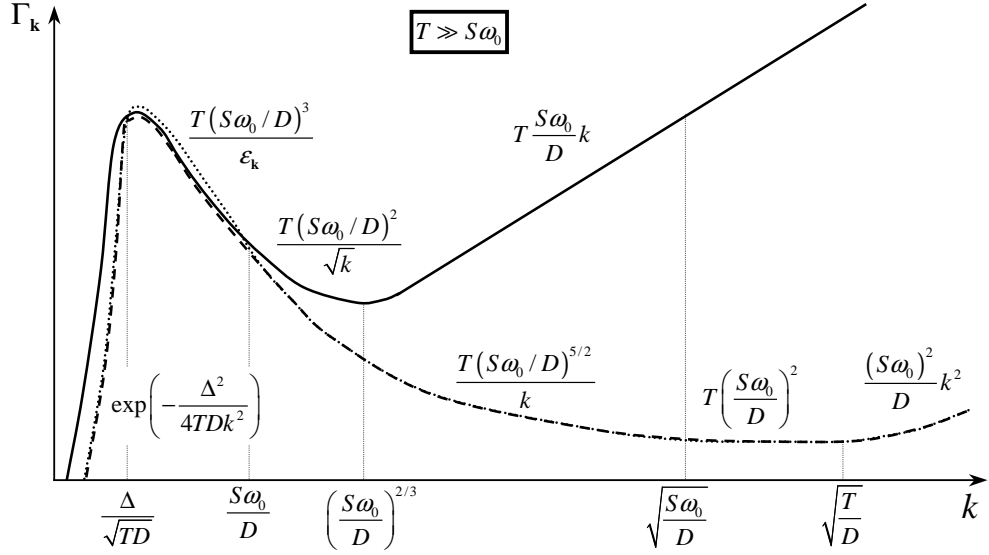}
\caption{Sketch of the spin-wave damping $\Gamma_{\bf k}$ versus the momentum at $k\ll1$ for $|\sin2\phi_{\bf k}|\sim1$ (solid line), $|\sin\phi_{\bf k}|=1$ (dashed line), and $|\sin\phi_{\bf k}|=0$ (dotted line). The corresponding dependences of $\Gamma_{\bf k}$ on $T$, $\omega_0$ and $k$ are also indicated. We imply $S\sim1$, and $T\gg S\omega_0$. Curves for $|\sin\phi_{\bf k}|=1$ and $|\sin\phi_{\bf k}|=0$ differ slightly only in the interval $\Delta/\sqrt{TD}\alt k \ll S\omega_0/D$. Notice that the ratio $\Gamma_{\bf k}/\epsilon_{\bf k}$ rises with decreasing $k$ for any given $\phi_{\bf k}$ at $k\agt\Delta/\sqrt{TD}$, there is the peak at $k\sim\Delta/\sqrt{TD}$ and the exponential decay at $k\alt\Delta/\sqrt{TD}$. The ratio $\Gamma_{\bf k}/\epsilon_{\bf k}$ is shown in the vicinity of the peak in Fig.~\ref{fdinf}.
\label{fdt}} 
\end{figure}

We sketch the dependence of $\Gamma_{\bf k}$ on the momentum in Fig.~\ref{fdt} at $k\ll 1$, $S\sim1$ and $T\gg S\omega_0$ taking into account the results obtained in this section. It is seen that the damping is highly anisotropic at $k\agt S\omega_0/D$. The damping increases with decreasing $k$ up to $k\sim\Delta/\sqrt{TD}$ if $\bf k$ is directed along a square edge (i.e., if $|\sin2\phi_{\bf k}|=0$). In contrast the damping is not monotonic function of $k$ for $|\sin2\phi_{\bf k}|\sim1$: it decreases with decreasing $k$ up to $k\sim(S\omega_0/D)^{2/3}$ and then it rises up to $k\sim\Delta/\sqrt{TD}$. The damping is only slightly anisotropic in the interval $\Delta/\sqrt{TD}\alt k\alt S\omega_0/D$. There is the peak at $k\sim\Delta/\sqrt{TD}$ at any given $\phi_{\bf k}$ which height can be estimated using Eq.~(\ref{damq<<}). This peak is followed by exponential decay of the damping at $k<\Delta/\sqrt{TD}$ having the form (\ref{damq<<<}).

To illustrate in more detail the region of small momenta $k\ll \sqrt{S\omega_0/D}$ in which the peak exists, we derive general expressions for $\Gamma_{\bf k}$ which coincide with Eq.~(\ref{damq<}) at $k\gg S\omega_0/D$ and with Eqs.~(\ref{damq<<}) and (\ref{ratio}) at $\Delta/\sqrt{TD}\alt k\ll S\omega_0/D$. These general expressions appear to be quite cumbersome for arbitrary $\phi_{\bf k}$. Then, we present here only the equation in the special case of $\sin\phi_{\bf k}=0$ which is the most simple one:
\begin{equation}
\label{rs}
\frac{\Gamma_{\bf k}}{\epsilon_{\bf k}} = 
\frac{\alpha^2 S^3 \omega_0^4}{16\pi D^2\Delta^2} 
\frac 1t
\frac{\sqrt{1+\kappa^2}}{\kappa^3}
\int_{\zeta}^\infty dq
\frac{\exp\left(q\sqrt{1+q^2}/t\right)}{\left(\exp\left(q\sqrt{1+q^2}/t\right) - 1\right)^2}
\frac{q(1+q^2)^{5/2}}{(1+2q^2)^3}\sqrt{1-\frac{1+q^2}{(1+2q^2)^2}\frac{1+\kappa^2}{\kappa^2}},
\end{equation}
where $\kappa=k\sqrt{SD\omega_0\alpha}/\Delta$, $t=T/(S\omega_0\alpha)$ and 
\begin{equation}
\label{zeta}
\zeta = \sqrt{\frac{1}{8\kappa^2}\left(1-3\kappa^2+\sqrt{9\kappa^4+10\kappa^2+1}\right)}. 
\end{equation}
At $q=\zeta$ the expression under the square root in Eq.~(\ref{rs}) is equal to zero and it is positive for $q>\zeta$. When $1\gg\kappa\agt\sqrt{S\omega_0\alpha/T}$ (i.e., when $\Delta/\sqrt{SD\omega_0}\gg k\agt\Delta/\sqrt{TD}$) and $T\gg S\omega_0$, Eq.~(\ref{rs}) transforms into Eq.~(\ref{damq<<}). In the opposite limiting case of $\kappa\gg1$ (to be precise, at $\Delta/\sqrt{SD\omega_0}\ll k \ll \sqrt{S\omega_0/D}$) one obtains the last term in Eq.~(\ref{damq<}) from Eq.~(\ref{rs}) at $T\gg S\omega_0$. 

We plot in Fig.~\ref{fdinf} the ratio of the spin-wave damping and the real part of the spectrum given by Eq.~(\ref{rs}) versus the reduced wave-vector $\kappa$ for 2D FM with $S=1/2$ and $S=3$ on the simple square lattice assuming that $\omega_0=0.01J$. The peak is seen at $\kappa\sim\sqrt{S\omega_0\alpha/T}$ (i.e., at $k\sim\Delta/\sqrt{TD}$). Its position moves to smaller $\kappa$ and the height rises as $S$ increases at a given ratio $T/T_C$ or as $T$ increases at a given $S$.

\section{Large spins}
\label{ls}

Let us consider now large spins $S\agt\ln \left(4\pi S (D/[S\omega_0])^{3/2}\right)$. As it is explained in the previous section, the temperature can be of the order of $D$ in this case remaining much smaller than the Curie temperature. When $T\sim D$ one can use expressions for $\Omega(\omega,{\bf k})$ obtained above with the exception for the gap: the temperature correction becomes important and one has for the gap in Eq.~(\ref{ot2}) 
\begin{equation}
\label{gapt}
\Delta_\gg^2 = 
\Delta^2 
+ 
\frac{S\omega_0\alpha}{\mathfrak N} 
\sum_{\bf q}
\frac{(E_{\bf q} - B_{\bf q}) N_{\bf q}}{\epsilon_{\bf q}}
\left[
\left(Q_{\bf q}^{xx} - Q_{\bf q}^{zz}\right)
- 
\frac{S (E_{\bf q} - B_{\bf q}) \left( Q_{\bf q}^{xz} \right)^2 }{\epsilon_{\bf q}^2}
\left(
1 + \frac{\epsilon_{\bf q}}{T} \left(1 + N_{\bf q}\right)
\right)
\right],
\end{equation}
where $\Delta$ is given by Eq.~(\ref{delta}), the first and the second terms in the square brackets stem from the Hartree-Fock and the loop diagrams shown in Fig.~\ref{diag}(b) and (c), respectively.

Let us discuss temperatures $2SJ_{\bf 0}<T\ll T_C^{(S\gg1)}$, where $2SJ_{\bf 0}$ is the spin-wave band width and $T_C^{(S\gg1)}$ is given by Eq.~(\ref{tci}). To find corrections to the spectrum at such large $T$ one can expand all Plank's functions under sums over momenta up to the first term: $N_{\bf q}\approx T/\epsilon_{\bf q}$. Then, we have Eq.~(\ref{ot2}) for the real part of $\Omega(\omega,{\bf k})$, where 
\begin{equation}
\label{renls}
W(T) = \frac{T}{SJ_{\bf 0}}, \quad V(T) = \frac{T}{2\pi D} \ln \left(\frac{D}{S\omega_0}\right)
\end{equation}
now and one obtains for the gap from Eq.~(\ref{gapt})
\begin{eqnarray}
\label{ginf}
\Delta_\gg &=& \sqrt{\Delta^2 + \alpha C_\gg S^2\frac{\omega_0^3}{D^2} T},\\
\label{cinf}
C_\gg &=& \frac{D^2}{\omega_0^2 \mathfrak N} \sum_{\bf q}\frac{\left( Q_{\bf q}^{xx} - Q_{\bf q}^{zz} \right)^2 - 4 \left( Q_{\bf q}^{xz} \right)^2 }{2\epsilon^2_{\bf q}},
\end{eqnarray}
where we can discard $\Delta^2$ under the square root in Eq.~(\ref{ginf}) because of its smallness compared to the second term at $T\gg D$. Summation over large momenta gives the main contribution in Eq.~(\ref{cinf}) and we find as a result of numerical computation for the coupling between only nearest neighbor spins on the simple square lattice $C_\gg \approx 0.025$.

A new wide region appears in the momentum space in which the bare spectrum is much larger than the gap and has the form $\epsilon_{\bf k}^2 \approx S\omega_0\alpha(Dk^2 + (S\omega_0k/2)\sin^2\phi_{\bf k})$. The corresponding interval in the $\bf k$-space is given by $(\Delta_\gg/(S\omega_0))^2\alt k \ll S\omega_0/D$ for $|\sin\phi_{\bf k}|\sim1$ and by $\Delta_\gg/\sqrt{SD\omega_0}\ll k \ll S\omega_0/D$ for $|\sin\phi_{\bf k}|\ll1$. As a result all the expressions for the damping obtained above should be reconsidered. In particular, Eq.~(\ref{damq>}) is valid only for $k\gg S\sqrt{\omega_0/T}$. All $T$-independent terms are negligible in Eq.~(\ref{damq>}) and one should use expression (\ref{ginf}) for the gap rather than Eq.~(\ref{delta}). Then, Eqs.~(\ref{damq>}) and (\ref{damq<}) are not valid at $\sqrt{S\omega_0/D}\ll k\ll S\sqrt{\omega_0/T}$ and $S\omega_0/D \ll k\ll\sqrt{S\omega_0/D}$, respectively, because integration over the above mentioned interval in the momentum space is essential in the corresponding sums. The results are quite cumbersome and we do not present them here. The main conclusion is that the spin-wave damping is much smaller than the real part of the spectrum at $k\gg S\omega_0/D$. Let us discuss in somewhat detail only the damping at small momenta $k\ll S\omega_0/D$ because the diffusion mode was proposed in Refs.~\cite{aprl,aprb} at small $k$.

The ratio of the spin-wave damping and the real part of the spectrum at 
$(\Delta_\gg/(S\omega_0))^2 \gg k \agt \Delta_\gg/\sqrt{TD}$ 
can be found from Eq.~(\ref{ratio}) (replacing $\Delta$ with $\Delta_\gg$) with the result
\begin{equation}
\label{ratiols}
\frac{\Gamma_{\bf k}}{\Delta_\gg} = \frac{1}{2^8C_\gg}.
\end{equation}
This expression gives the approximate value of the peak height on the curve $\Gamma_{\bf k}/\epsilon_{\bf k}$ that is valid for all $\phi_{\bf k}$. The right part of Eq.~(\ref{ratiols}) is equal approximately to $0.16$ for the exchange coupling between only nearest neighbor spins on the simple square lattice. It is interesting to note that the smallness of the spin-wave damping in this case is {\it numerical} whereas at $S\sim1$ the smallness is parametric (see Eq.~(\ref{ratio})). Notice that at $k\ll\Delta_\gg/\sqrt{TD}$ the damping is exponentially small, as it is discussed in Sec.~\ref{ksm}. 

\begin{figure}
\centering
\includegraphics{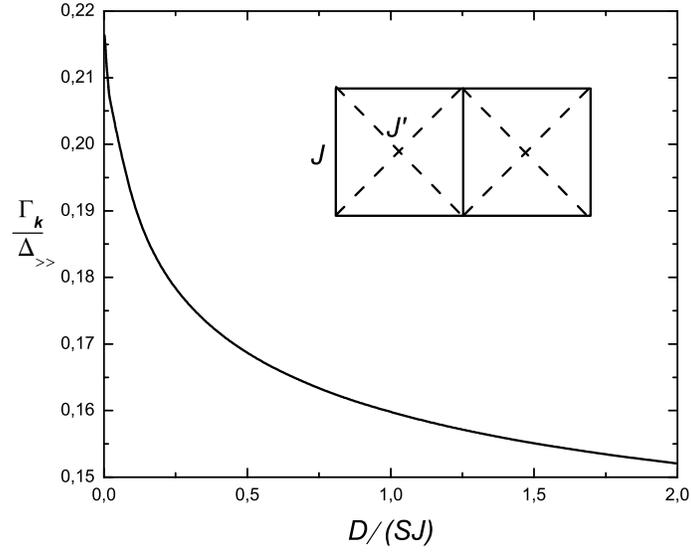}
\caption{The peak height on the curve for $\Gamma_{\bf k}/\epsilon_{\bf k}$ given by Eq.~(\ref{ratiols}) versus the dimensionless spin-wave stiffness $D/(SJ)$ for 2D FM with $S\gg1$ and with the next-nearest-neighbor exchange coupling $J'$ (see the inset). One has $D=SJ(1+2J'/J)$ in this case and it is implied that $D\gg S\omega_0$ and $8S(J+J')\ll T\ll T_C\propto SD$ even for small $D$, where $8S(J+J')$ is the spin-wave band width.
\label{fdfrust}} 
\end{figure}

It is seen from Eq.~(\ref{ratiols}) that the value of the peak height is in inverse proportion to $C_\gg$ which depends on the exchange coupling between spins $J_{lm}$ (see Eq.~(\ref{cinf})). It is interesting to examine the dependence of $\Gamma_{\bf k}/\Delta_\gg$ given by Eq.~(\ref{ratiols}) on the value of the coupling between next-nearest neighbors which, in particular, can reduce $D$ significantly. We assume that the exchange coupling between nearest- and next-nearest neighbor spins are equal to $J$ and $J'$, respectively (see the inset in Fig.~\ref{fdfrust}). One has for the spin-wave stiffness in this case $D=SJ(1+2J'/J)$. The dependence of the peak height on $D/(SJ)$ is shown in Fig.~\ref{fdfrust}. It is seen that even at very small $D$, i.e., for frustrating next-nearest-neighbor interaction, magnons are well-defined quasi-particles. Importantly, it is implied even in the case of small $D$ that $D\gg S\omega_0$ and $8S(J+J')\ll T\ll T_C^{(S\gg1)}\propto SD$, where $8S(J+J')$ is the spin-wave band width.

We do not present here expressions for the damping at $(\Delta_\gg/(S\omega_0))^2\alt k\alt S\omega_0/D$ for arbitrary $\phi_{\bf k}$. The main conclusion is that the damping is much smaller than the real part of the spectrum. To illustrate this let us consider only $\sin\phi_{\bf k}=0$. The bare spectrum has the minimum value at $\sin\phi_{\bf k}=0$ for a given $k$ in the discussed momentum interval and it is the most "dangerous" case in which one could expect large spin-wave damping as compared with the real part of the spectrum. Corresponding calculations give Eq.~(\ref{rs}) for the damping at $k\ll \sqrt{S\omega_0/D}$, where one should replace $\Delta$ with $\Delta_\gg$. We use this modification of Eq.~(\ref{rs}) to plot $\Gamma_{\bf k}/\epsilon_{\bf k}$ in Fig.~\ref{fdinf} for $S=30$, $\omega_0=0.01J$ and $T=0.1T_C^{(S\gg1)}$.

\subsection*{Classical spins}

To find the spectrum renormalization in the classical 2D FM using the spin-wave formalism discussed in the present paper one should consider the limit of 
\begin{equation}
\label{lim}
S\to\infty, \quad \hbar\to0, \quad J,\omega_0\to0
\end{equation}
assuming that 
\begin{equation}
\label{ass}
\hbar S = {\rm const}, \quad JS^2 = j = {\rm const}, \quad \omega_0S^2 = w = {\rm const}
\end{equation}
and $T/j$ is much smaller than unity. Moreover, one should replace operators of creation and annihilation $a_{\bf k},a_{\bf k}^\dagger$ with classical operators $\beta_{\bf k}=a_{\bf k}/\sqrt S$, $\beta_{\bf k}^\dagger=a_{\bf k}^\dagger/\sqrt S$ which Bose occupation numbers are finite. \cite{har,loly} As a result the spectrum and corrections to it in the classical limit can be obtained from the expressions found above by multiplying them by $S$ and taking the limit (\ref{lim}) with the assumptions (\ref{ass}). In particular, we have for the gap in the classical 2D FM from Eq.~(\ref{ginf})
\begin{equation}
\label{ginf2}
\Delta_\infty = \sqrt{\alpha C_\gg \frac{w^3}{j^2} T}.
\end{equation}
Only the second term under the square root in Eq.~(\ref{ginf}) contributes to $\Delta_\infty^2$. Notice that there is no exponential decay of the damping at small $k$ in classical 2D FM in which Eq.~(\ref{ratiols}) is valid at all $k$ much smaller than $C_\gg  wT/j^2$ because $\Delta_\gg/\sqrt{TD}\to0$ when $S\to\infty$. Then, the peak in $\Gamma_{\bf k}$ and $\Gamma_{\bf k}/\epsilon_{\bf k}$ is located at $k=0$.
 
\begin{figure}
\centering
\includegraphics{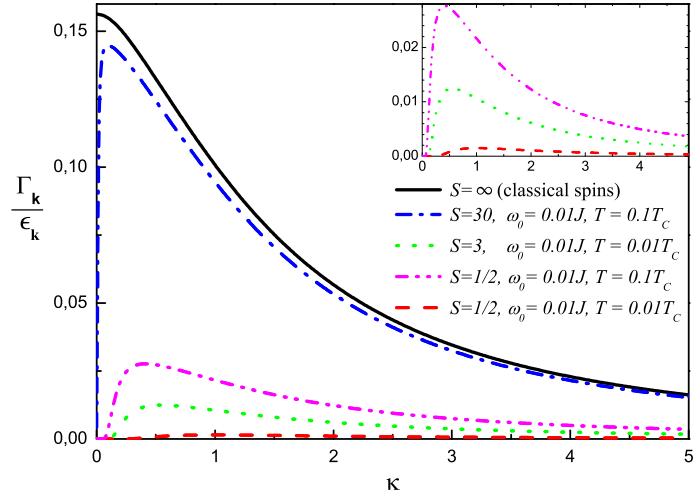}
\caption{(Color online.) The ratio of the spin-wave damping and the real part of the spectrum $\Gamma_{\bf k}/\epsilon_{\bf k}$ versus the reduced wave-vector $\kappa$ at $\sin\phi_{\bf k}=0$ for the classical and quantum 2D FMs on the simple square lattice. The ratios $\Gamma_{\bf k}/\epsilon_{\bf k}$ are given by Eqs.~(\ref{rinf}) and (\ref{rs}) in which $\kappa=k\sqrt{ j^3/(C_\gg w^2T)}$ and $\kappa=k\sqrt{SD\omega_0\alpha}/\Delta_\gg$ for the classical and quantum FMs, respectively, where $\Delta_\gg$ is given by Eq.~(\ref{gapt}). The curve for classical spins is for $T\ll j$ and $w\ll j$ (see Eqs.~(\ref{ass})). There are peaks at $\kappa\sim\sqrt{S\omega_0\alpha/T}$ in quantum 2D FMs. Inset: the same is shown for $S=1/2$ and $S=3$ on a large scale. For all quantum magnets the boundary of the area shown in Fig.~\ref{damp} in which the spin-wave damping is zero at $T=0$ is located at $\kappa\agt 10$.
\label{fdinf}} 
\end{figure}

In particular, one obtains in the limiting case of classical spins from Eq.~(\ref{rs}) by expanding the exponents in $t$ and replacing $\Delta$ with $\Delta_\infty$
\begin{equation}
\label{rinf}
\frac{\Gamma_{\bf k}}{\epsilon_{\bf k}} = \frac{1}{16\pi C_\gg}
\frac{\sqrt{1+\kappa^2}}{\kappa^3}
\int_{\zeta}^\infty dq\frac{(1+q^2)^{3/2}}{q(1+2q^2)^3}\sqrt{1-\frac{1+q^2}{(1+2q^2)^2}\frac{1+\kappa^2}{\kappa^2}},
\end{equation}
where $\kappa=k \sqrt{j^3/(C_\gg w^2T)}$ now and $\zeta$ is given by Eq.~(\ref{zeta}). Eq.~(\ref{rinf}) transforms into Eq.~(\ref{ratiols}) at $\kappa\ll1$ (with $\Delta_\infty$ put instead of $\Delta_\gg$). We plot in Fig.~\ref{fdinf} also the ratio of the spin-wave damping and the real part of the spectrum given by Eq.~(\ref{rinf}) versus the reduced wave-vector $\kappa$ for the classical 2D FM on the simple square lattice. As it is pointed out above, the position of the peak on quantum curves moves to smaller $\kappa$ and the peak height rises as $S$ increases at a given ratio $T/T_C$ or as $T$ increases at a given $S$. But as it is demonstrated above and as it is seen by the example of $S=1/2$, $S=3$ and $S=30$ shown in Fig.~\ref{fdinf}, the peaks on the quantum curves cannot be higher than that of the classical 2D FM located at $k=0$ which height is given by Eq.~(\ref{ratiols}). It should be noted here that the peak height for finite $S\gg1$ and $D\ll T\ll T_C^{(S\gg1)}$ is slightly smaller than that given by Eq.~(\ref{ratiols}) because we have discarded $\Delta^2$ under the square root in Eq.~(\ref{ginf}) deriving Eq.~(\ref{ratiols}). On the other hand Eq.~(\ref{ratiols}) gives the precise value of the peak height for classical 2D FM because $\Delta^2$ disappears after taking the limit (\ref{lim}) with the assumptions (\ref{ass}) and only the second term under the square root in Eq.~(\ref{ginf}) contributes to the gap.

\section{Discussion}
\label{disc}

We address in this section five questions: (i) relation between the anisotropic term (\ref{an}) in the total energy of 2D FM and the spin-wave gap (\ref{delta}), (ii) discussion of the further order $1/S$ corrections, (iii) calculation of the magnetization taking into account the spin-wave spectrum renormalization, (iv) derivation of the spin Green's functions in the first order of $1/S$, and (v) brief discussion of the effect of the easy-plane anisotropy and the spectrum renormalization obtained in Refs.~\cite{aprl,aprb}.

(i) 
As is discussed in Introduction, it looks reasonable that the dipolar in-plain anisotropy given by Eq.~(\ref{an}) should be accompanied with a spin-wave gap. The relation between this anisotropy and the gap can be shown for 2D FM in the same non-rigorous manner as for 3D FM. \cite{i} Let us discuss large spins and try to take into account the dipolar in-plain anisotropy (\ref{an}) phenomenologically by adding to the microscopic Hamiltonian (\ref{ham0}) the following expression [cf. Eq.~(\ref{an})]:
\begin{equation}
\label{anphen}
C\frac{\omega_0^2}{2S^2D}\sum_l \left(S_l^x\right)^2\left(S_l^z\right)^2 .
\end{equation}
This term after Dyson-Maleev transformation (\ref{dm}) gives the contribution $CS\omega_0^2/(4D)(a^\dagger_{\bf k} + a_{\bf k})^2$ to the bilinear part (\ref{h2}) of the Hamiltonian that in turn leads to the shift $E_{\bf k}\mapsto E_{\bf k} + CS\omega_0^2/(2D)$ and $B_{\bf k}\mapsto B_{\bf k} + CS\omega_0^2/(2D)$. Using this renormalization of $E_{\bf k}$ and $B_{\bf k}$ and Eq.~(\ref{spec1}) for the spectrum one recovers the spin-wave gap (\ref{delta}) in Eq.~(\ref{ot1}). This consideration does not work for $S=1/2$ because Eq.~(\ref{anphen}) is a constant in this case.

\begin{figure}
\centering
\includegraphics{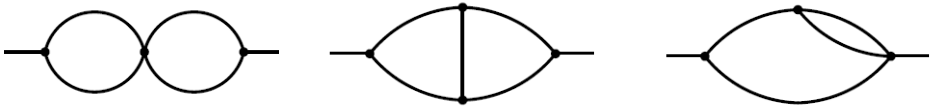}
\caption{Some diagrams of the second order in $1/S$.
\label{so}} 
\end{figure}

(ii) 
Let us turn to further order $1/S$ corrections. Some diagrams of the second order in $1/S$ are presented in Fig.~\ref{so}. It can be shown that at least their real parts are much smaller than the real parts of the first order diagrams discussed above. To demonstrate this one has to take into account that the spin-wave gap screens infrared singularities appearing in some of these diagrams. Moreover three- and four-particle vertexes contain additional smallnesses at small external momenta. Thus, the three-particle vertex (\ref{h3}) contains $xz$-component of the dipolar tensor which is proportional to the product of $\omega_0$ and momentum (see Eq.~(\ref{qsmall})). As for four-particles vertex, the expression under the sum in Eq.~(\ref{h4}) has the form at $k_{1,2,3,4}\ll1$
\begin{eqnarray}
\label{h4s}
&&2{\bf k}_3(2{\bf k}_1+{\bf k}_3)a^\dagger_{-1}a^\dagger_{-2}a_3a_4 
- \nonumber\\
&& \omega_0k_2\frac{1+\sin^2\phi_{{\bf k}_2}}{2} a^\dagger_{-1} a_2a_3a_4
+\\
&& \frac{\omega_0}{2} \left(k_2\sin^2\phi_{{\bf k}_2} + 2\left|{\bf k}_2+{\bf k}_3\right|\cos^2\phi_{{\bf k}_2+{\bf k}_3}\right)  a^\dagger_{-1} a^\dagger_{-2} a_3a_4
+\nonumber\\
&& \alpha\omega_0 a^\dagger_{-1}\left(a_2-a^\dagger_{-2}\right)a_3a_4.\nonumber
\end{eqnarray}
It is seen that the first term in Eq.~(\ref{h4s}) is quadratic in momenta. The second and the third ones are proportional to $\omega_0$ and momenta. The last term in Eq.~(\ref{h4s}) is proportional only to $\omega_0$ but the combination $a_2-a^\dagger_{-2}$ involved in it is "soft": for instance, it's coupling with operator $a_{-2}$ gives $F(\omega_2,{\bf k}_2)-\overline{G}(\omega_2,{\bf k}_2)\approx -(Dk_2^2+S\omega_0k_2\sin^2\phi_{{\bf k}_2}/2-\omega_2)/{\cal D}(\omega_2,{\bf k}_2)$ that is much smaller than $F(\omega_2,{\bf k}_2)$ or $\overline{G}(\omega_2,{\bf k}_2)$ themselves ($F(\omega_2,{\bf k}_2)\approx G(\omega_2,{\bf k}_2)\sim S\omega_0/{\cal D}(\omega_2,{\bf k}_2)$ at $k_2\ll\sqrt{S\omega_0/D}$ and $\omega_2\ll S\omega_0$). As a result further order diagrams appear to be small at $T\ll T_C$.

Unfortunately, the smallness of the three- and four-particle vertexes was not taken into account in my previous paper \cite{i} devoted to 3D FM in similar qualitative discussion of further order $1/S$ corrections. As a result further order corrections were overestimated there. It was proposed that they are small only at $T\ll\omega_0$, whereas the range of the validity of the perturbation theory is much wider in 3D FM: $T\ll T_C$. 

\begin{figure}
\centering
\includegraphics{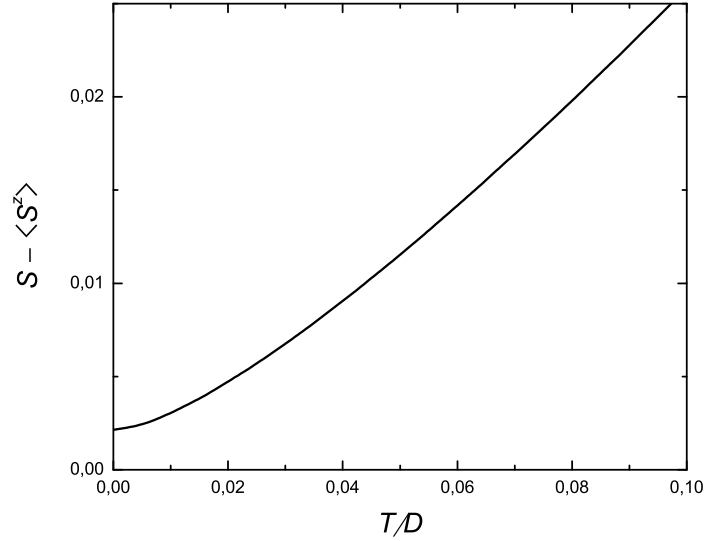}
\caption{The value $S - \langle S^z\rangle$ given by Eq.~(\ref{mag}) versus the dimensionless temperature $T/D$ for the square 2D FM with $\omega_0=0.1J$.
\label{magfig}} 
\end{figure}

(iii) 
Let us calculate now the magnetization value $\langle S^z\rangle$ using renormalized spectrum. We have after simple computation
\begin{eqnarray}
\label{mag}
\frac{S - \langle S^z\rangle}{S} =
\frac{1}{\mathfrak N} \sum_{\bf q} \frac{E_{\bf q}(1+2N_{\bf q}) - \epsilon_{\bf q}}{2S\epsilon_{\bf q}} 
= \left\{
\begin{array}{ll}
\displaystyle \frac{T}{4\pi DS} \ln \left(\frac{T}{D} \left(\frac{D}{S\omega_0}\right)^{3/2}\right), & \displaystyle T\gg \frac{S\omega_0}{\ln (D/[S\omega_0])},\\
\displaystyle \frac{\omega_0\alpha}{16\pi D}, & \displaystyle T\ll \frac{S\omega_0}{\ln (D/[S\omega_0])}.
\end{array}
\right.
\end{eqnarray}
Notice that zero-point fluctuations give the main contribution at $T\ll S\omega_0/\ln (D/[S\omega_0])$. The value $S - \langle S^z\rangle$ given by Eq.~(\ref{mag}) is shown in Fig.~\ref{magfig} for $\omega_0=0.1J$. It should be pointed out also that taking into consideration the gap in the spectrum does not change the form of the magnetization. Meantime Eq.~(\ref{mag}) differs from the corresponding expression in Ref.~\cite{mal}. The origin of this discrepancy is discussed in Appendix~\ref{apmag}. The value of the Curie temperature (\ref{tc}) giving by the spin-wave theory follows from Eq.~(\ref{mag}) by putting $\langle S^z\rangle$ to be equal to zero. In the case of large spins $S\agt\ln \left(4\pi S (D/[S\omega_0])^{3/2}\right)$ the Plank's function in Eq.~(\ref{mag}) can be expanded and one finds Eq.~(\ref{tci}) for the Curie temperature by putting $\langle S^z\rangle=0$.

(iv) 
Spin Green's functions defined as 
$\chi_{ij}(\omega,{\bf k}) = i\int_0^\infty dt e^{i\omega t} \langle [S^i_{-\bf k}(t), S^j_{\bf k}(0)] \rangle$ 
can be calculated straightforwardly in the first order of $1/S$ using Eqs.~(\ref{gf})--(\ref{o}) and expressions for the self-energy parts with the following results for the transverse components
\begin{eqnarray}
\label{xx}
\chi_{xx}(\omega,{\bf k}) &=& -S \frac{Dk^2 + S\omega_0\alpha}{\omega^2 - \epsilon_{\bf k}^2 - \Omega(\omega,{\bf k})},\\
\label{yy}
\chi_{yy}(\omega,{\bf k}) &=& -S \frac{Dk^2 + (S\omega_0/2)k\sin^2\phi_{\bf k} + \Delta^2/(S\omega_0\alpha)}{\omega^2 - \epsilon_{\bf k}^2 - \Omega(\omega,{\bf k})},
\end{eqnarray}
where $\epsilon_{\bf k}$ is the bare spectrum here. Corrections from the self-energy parts to numerator are negligible in Eq.~(\ref{xx}). In contrast numerator in Eq.~(\ref{yy}) for $\chi_{yy}(\omega,{\bf k})$ renormalizes greatly at small $k$ so that the uniform susceptibility $\chi_{yy}(\omega,{\bf 0})$ becomes finite. 

The corresponding expression for the longitudinal component $\chi_{zz}(\omega,{\bf k})$ is slightly more cumbersome. It can be calculated using Eqs.~(\ref{imsum}). Let us discuss only uniform longitudinal susceptibility $\chi_{zz}(\omega,{\bf 0})$. We have for $|\omega| \gg \sqrt{(S\omega_0)^3/D} \gg \Delta$
\begin{equation}
\label{zz1}
\chi_{zz}(\omega,{\bf 0}) = \frac{TS\omega_0\alpha}{4D}\frac{1}{\omega^2}
\left(
\frac2\pi \ln\left|\frac{2\Delta}{\omega}\right|
+
i{\rm sgn}(\omega)
\right).
\end{equation}
One obtains for the imaginary part when $|\omega| \alt \sqrt{(S\omega_0)^3/D}$
\begin{equation}
\label{zz2}
{\rm Im}\chi_{zz}(\omega,{\bf 0}) = 
{\rm sgn}(\omega)\theta(|\omega|-2\Delta) 
\frac{\alpha^{3/4}\Gamma(3/4)^2}{\sqrt{2\pi^3}}
\frac{T(S\omega_0)^{1/4}}{D^{3/4}}
\frac{(\omega^2 - 4\Delta^2)^{1/4}}{\omega^2},
\end{equation}
where $\Gamma(x)$ is the gamma-function and $\theta(x)$ is the theta-function ($\theta(x)=1$ when $x>0$ and $\theta(x)=0$ when $x<0$). The corresponding expression for the real part of $\chi_{zz}(\omega,{\bf 0})$ is quit cumbersome and we do not present it here. It is seen from Eq.~(\ref{zz2}) that ${\rm Im}\chi_{zz}(\omega,{\bf 0})=0$ at $\omega<2\Delta$. In contrast, if one did not take into account the spin-wave gap $\Delta$ the infrared singularity would appear of the form ${\rm Im}\chi_{zz}(\omega\to0,{\bf 0})\sim\omega^{-3/2}$. Such a singularity is nonphysical one since it leads to an infinitely large absorption function \cite{landau_sp} $Q_\omega\propto\omega{\rm Im}\chi(\omega)$ at $\omega=0$. Similar situation exists in 3D FM with dipolar forces. The infrared singularity of the form ${\rm Im}\chi_{zz}(\omega\to0,{\bf 0}) \sim T/\omega$ can be found not taking into account the spin-wave gap. \cite{top} This singularity leads to the finite absorption function at $\omega=0$ signifying that the sample would be heated by a dc field. On the other hand the spin-wave gap screens this singularity leading to zeroth $Q_\omega$ at $\omega=0$. \cite{i}

(v) 
We will discuss in detail elsewhere the effect of the exchange anisotropy having the form 
\begin{equation}
\label{ang}
{\cal H}_a = \frac12\sum_{l\ne m} A_{lm} S_l^y  S_m^y.
\end{equation}
and the one-ion anisotropy $\sum_l A(S_l^y)^2$. The corresponding renormalization of the above results for the real part of the spectrum is discussed briefly in Appendix~\ref{anis} taking into account the exchange anisotropy (\ref{ang}). In particular, it is shown there that expressions (\ref{delta}) and (\ref{ginf}) for the gap should be multiplied by the factor $\sqrt{\tilde\omega_0/\omega_0}$, where
\begin{equation}
\label{ot}
\tilde\omega_0 = \omega_0 + \frac{A_{\bf 0}}{\alpha}.
\end{equation}
In the limiting case of classical spins one should imply in addition to Eq.~(\ref{lim}) and Eq.~(\ref{ass}) that $A\to0$ and $AS^2={\rm const}$. Then, we will assume that $\tilde\omega_0S^2=\tilde w={\rm const}$ and one has for the gap in classical 2D FM Eq.~(\ref{ginf2}) multiplied by $\sqrt{\tilde w/w}$.

Classical 2D FM with dipolar interaction and easy-plane anisotropy was discussed in Refs.~\cite{aprl,aprb}. 
Existence of the easy-plane anisotropy seems to be not crucial for the results obtained in Refs.~\cite{aprl,aprb}. The condition $\sqrt{\tilde w/j} \gg w/j$, that is important for the consideration in these papers, holds also at $A=0$ ($\tilde w = w$) if $w\ll j$ that is also implied there.
The spin-wave gap (\ref{ginf2}) (multiplied by $\sqrt{\tilde w/w}$) was not taken into account in Refs.~\cite{aprl,aprb}. Great renormalization of the spin-wave spectrum at small enough momenta was obtained there. In particular, a diffusion mode was found at 
$k\ll k_{DM}\sim wt^{3/4}/[j\ln^{1/4}(\sqrt{j\tilde w/w})]$, 
where $t=T/(4\pi j)$. Meantime the spin-wave gap screens all the spectrum peculiarities obtained in Refs.~\cite{aprl,aprb}. For example, at $|\sin\phi_{\bf k}|\ll1$ the energy of the diffusion mode has the form up to a numerical factor of the order of unity
$-ik^2 t^{-1/4}\ln^{5/4}(\sqrt{j\tilde w/w})\sqrt{\tilde wj^3/w^2}$. The spin-wave gap given by Eq.~(\ref{ginf2}) multiplied by $\sqrt{\tilde w/w}$ is much larger than the energy of the diffusion mode at $k\ll k_{DM}$.

\section{Summary and conclusion}
\label{con}

In the present paper we discuss two-dimensional Heisenberg ferromagnet (2D FM) with dipolar forces at $0\le T\ll T_C$ described by the Hamiltonian (\ref{ham0}) and consider renormalization of the bare spin-wave spectrum (\ref{spec1}) due to interaction between spin waves. For this purpose we carry out a comprehensive analysis of the first $1/S$ corrections to the spin-wave spectrum originating from diagrams shown in Fig.~\ref{diag}. 

We obtain the following results for $S\sim1$. Corrections to the square of the real part of the spectrum are given by Eqs.~(\ref{ot1}) and (\ref{ot2}). In particular, at $T\gg S\omega_0$, where $\omega_0$ is the characteristic dipolar energy given by Eq.~(\ref{o0}), thermal corrections result simply in renormalization (\ref{ren}) of the constants $D$ (spin-wave stiffness) and $\omega_0$ in the bare spectrum (\ref{spec1}). This renormalization is small at $T\ll T_C$. But it is significant that similar to 3D FM considered in our previous paper \cite{i} we obtain also the spin-wave gap $\Delta$ in the spectrum given by Eq.~(\ref{delta}). This gap stemming from the spin-wave interaction renormalizes greatly the bare gapless spectrum (\ref{spec1}) at $k\alt(\Delta/(S\omega_0))^2$ (see Fig.~\ref{spec2d}). 

Spin-wave damping $\Gamma_{\bf k}$ at $T=0$ is given by Eqs.~(\ref{damq>}) and (\ref{damq<}) for momenta $1\gg k\gg\sqrt{S\omega_0/D}$ and $k\ll\sqrt{S\omega_0/D}$, respectively. There is a region in the $\bf k$ space at small momenta sketched in Fig.~\ref{damp} in which the damping is equal to zero. Meantime we find great thermal enhancement of the damping in this region.

Temperature fluctuations give the main contribution to the spin-wave damping at $T\gg S\omega_0$ that is given by Eqs.~(\ref{damq>}), (\ref{damq<}), (\ref{damq<<}), and (\ref{damq<<<}) for momenta $k\gg\sqrt{S\omega_0/D}$, $S\omega_0/D\ll k\ll\sqrt{S\omega_0/D}$, $\Delta/\sqrt{TD}\alt k \ll S\omega_0/D$, and $k < \Delta/\sqrt{TD}$, respectively. The dependence of the damping on the momentum is sketched in Fig.~\ref{fdt} for $k\ll1$. It is seen that $\Gamma_{\bf k}$ is highly anisotropic at $k\agt S\omega_0/D$. The damping increases with decreasing $k$ up to $k\sim\Delta/\sqrt{TD}$ if $\bf k$ is directed along a square edge (i.e., if $|\sin2\phi_{\bf k}|=0$). In contrast the damping is not monotonic function of $k$ when $|\sin2\phi_{\bf k}|\sim1$: it decreases with decreasing $k$ up to $k\sim(S\omega_0/D)^{2/3}$ and then it rises up to $k\sim\Delta/\sqrt{TD}$. The damping is only slightly anisotropic in the interval $\Delta/\sqrt{TD}\alt k\alt S\omega_0/D$. There is a peak at $k\sim\Delta/\sqrt{TD}$ at any given $\phi_{\bf k}$ which height can be estimated using Eq.~(\ref{damq<<}). If the temperature is large enough so that the interval is finite given by $(\Delta/(S\omega_0))^2\gg k\agt\Delta/\sqrt{TD}$, the peak height is given by Eq.~(\ref{ratio}). This peak is followed by exponential decay of the damping at $k<\Delta/\sqrt{TD}$ having the form (\ref{damq<<<}).

An important quantity to be examined is the ratio of the spin-wave damping and the real part of the spectrum $\Gamma_{\bf k}/\epsilon_{\bf k}$ which is much smaller than unity if magnons are well-defined quasi-particles. It is seen from Eqs.~(\ref{damq>}), (\ref{damq<}), and (\ref{damq<<}) that the ratio $\Gamma_{\bf k}/\epsilon_{\bf k}$ rises upon decreasing $k$ at $k\agt S\omega_0/D$ for all $\phi_{\bf k}$. This growth changes into exponential decay at $k\alt\Delta/\sqrt{TD}$ given by Eq.~(\ref{damq<<<}). Then, there is a peak in $\Gamma_{\bf k}/\epsilon_{\bf k}$ at $k\sim\Delta/\sqrt{TD}$ and at any given $\phi_{\bf k}$. This peak rises and its position moves to smaller $k$ as $S$ increases at a given ratio $T/T_C$ or as $T$ increases at a given $S$ (see Fig.~\ref{fdinf}). Meantime its height is restricted by Eq.~(\ref{ratio}) which is proportional to $T/D$. In the case of $S\sim1$ we have $T/D\ll1$ when $T\ll T_C$. 

On the other hand, as it is pointed out in Sec.~\ref{damping}~C, the temperature can be greater than $D$ for large spin values much greater than unity, $S\agt\ln \left(4\pi S (D/[S\omega_0])^{3/2}\right)$, so that $D< T\ll T_C\propto DS$. Renormalization of the spectrum should be reconsidered at such large $T$ and $S$ because new large temperature $1/S$ corrections arise. Such reconsideration is carried out in Sec.~\ref{ls}. In particular, we find thermal renormalization (\ref{ren}) of the constants $D$ and $\omega_0$ in the bare spectrum (\ref{spec1}), where $W(T)$ and $V(T)$ are given by Eqs.~(\ref{renls}) in this case. Then, thermal correction to the gap becomes important and we have Eq.~(\ref{gapt}) for the gap at $T\sim D$ which transform into Eq.~(\ref{ginf}) at $2SJ_{\bf 0}\ll T\ll T_C$, where $2SJ_{\bf 0}$ is the spin-wave band width. Focusing on the spin-wave damping renormalization at small $k$ only, we observe that the peak height of $\Gamma_{\bf k}/\epsilon_{\bf k}$ cannot exceed the value given by Eq.~(\ref{ratiols}) that is equal approximately to 0.16 for the simple square lattice and that is a counterpart of Eq.~(\ref{ratio}) for $S\sim1$. It is interesting to note the {\it numerical} smallness of the peak height in $\Gamma_{\bf k}/\epsilon_{\bf k}$ when $S\gg1$ and $T_C\gg T\gg 2SJ_{\bf 0}$ that contrasts to the case of small $S$, in which the peak in $\Gamma_{\bf k}/\epsilon_{\bf k}$ cannot exceed the value (\ref{ratio}) proportional to $T/D\ll1$. The small value given by Eq.~(\ref{ratiols}) increases only slightly upon taking into account a frustrating next-nearest-neighbor exchange coupling (see Fig.~\ref{fdfrust}). The limiting case of classical spins is also discussed in Sec.~\ref{ls} at $T\ll j$, where $j$ is the exchange constant in the classical model (see Eq.~(\ref{ass})). We obtain expression (\ref{ginf2}) for the gap and the peak in $\Gamma_{\bf k}/\epsilon_{\bf k}$ at $k=0$ which height is given by Eq.~(\ref{ratiols}) {\it precisely} (see Fig.~\ref{fdinf}). {\it Thus, we find that magnons are well-defined quasi-particles in both quantum and classical 2D FMs with dipolar forces.}

We note that appearance of the gap is accompanied by the anisotropy in the total energy of the quantum 2D FM given by Eq.~(\ref{an}) and caused by quantum fluctuations lifting the degeneracy of the classical ground state. We demonstrate in Sec.~\ref{disc} the relation between the dipolar anisotropic term (\ref{an}) in the total energy and the gap in the spectrum at $T=0$.

Spin Green's functions $\chi_{ij}(\omega,{\bf k})$ are derived in Sec.~\ref{disc} with the results (\ref{xx}) and (\ref{yy}) for the transverse diagonal components $\chi_{xx}(\omega,{\bf k})$ and $\chi_{yy}(\omega,{\bf k})$, and (\ref{zz1}) and (\ref{zz2}) for uniform longitudinal one $\chi_{zz}(\omega,{\bf 0})$. It should be noted that if one did not take into account the spin-wave gap the infrared singularity would appear of the form ${\rm Im}\chi_{zz}(\omega\to0,{\bf 0})\sim\omega^{-3/2}$. Such a singularity is nonphysical one since it leads to an infinitely large absorption function $Q_\omega\propto\omega{\rm Im}\chi(\omega)$ at $\omega=0$. The spin-wave gap screens this singularity: as it is seen from Eq.~(\ref{zz2}), ${\rm Im}\chi_{zz}(\omega,{\bf 0})=0$ at $\omega<2\Delta$. 

Modification of the results by taking into consideration the easy-plane exchange anisotropy (\ref{ang}) is discussed briefly in Sec.~\ref{disc}. In particular, we have shown that expressions (\ref{delta}) and (\ref{ginf}) for the gap should be multiplied by the factor $\sqrt{\tilde\omega_0/\omega_0}$, where $\tilde\omega_0$ is given by Eq.~(\ref{ot}).

Expression (\ref{mag}) for the magnetization in 2D FM is obtained which differs from the well-known result of Ref.~\cite{mal}. Higher order corrections to the spectrum are discussed in Sec.~\ref{disc} and it is concluded that they are small compared to the first corrections obtained. 

We would like to note in conclusion that the spectrum is gapless in 3D antiferromagnets with dipolar forces in the spin-wave approximation \cite{3daf} and the spin-wave interaction should lead to the gap in 3D antiferromagnets similar to 2D and 3D FMs.

\begin{acknowledgments}

This work was supported by Russian Science Support Foundation, President of Russian Federation (grant MK-1056.2008.2), RFBR grant 06-02-16702, and Russian Programs "Quantum Macrophysics", "Strongly correlated electrons in semiconductors, metals, superconductors and magnetic materials" and "Neutron Research of Solids".

\end{acknowledgments}

\appendix

\section{Calculation of $\Omega^{(3)}(\omega,{\bf k})$}
\label{loop}

We present in this appendix some details of calculation of $\Omega^{(3)}(\omega,{\bf k})$ that is a contribution to $\Omega(\omega,{\bf k})$ given by Eq.~(\ref{o}) from the loop diagram shown in Fig.~\ref{diag}(c). This diagram originates from ${\cal H}_3$ terms (\ref{h3}) in the Hamiltonian. As a result of simple calculations we lead to quit a cumbersome expression
\begin{subequations}
\label{o3}
\begin{eqnarray}
\Omega^{(3)}(i\omega,{\bf k}) &=& 
-Dk^2 \frac{S}{\mathfrak N} T\sum_{q_1 + q_2 = k} \frac{1}{[(i\omega_1)^2 - \epsilon_1^2][(i\omega_2)^2 - \epsilon_2^2]} \nonumber\\
&&\times\Bigl( \left( Q_{\bf k}^{xz}\right)^2 (B_1B_2 + E_1E_2 + \omega_1\omega_2) \\
&&{}
+ 2Q_{\bf k}^{xz}Q_1^{xz} ([E_1-B_1][E_2-B_2] + \omega_1\omega_2) \\
&&{}
+ Q_1^{xz}Q_2^{xz} ([E_1-B_1][E_2-B_2] - \omega_1\omega_2) + 2\left( Q_2^{xz}\right)^2 E_1(E_2 - B_2) \Bigr) \\
&&{}
-\frac{S^2\omega_0\alpha}{\mathfrak N} T\sum_{q_1 + q_2 = k} \frac{1}{[(i\omega_1)^2 - \epsilon_1^2][(i\omega_2)^2 - \epsilon_2^2]} \nonumber\\
&&\times\Bigl(   Q_1^{xz}(Q_1^{xz} + Q_2^{xz}) (E_1-B_1)(E_2-B_2) \\
&&{}
+ \left(Q_{\bf k}^{xz}\right)^2 (B_1B_2 + E_1E_2 + \omega_1\omega_2) \\
&&{}+ 2Q_{\bf k}^{xz}Q_1^{xz} ([E_1-B_1][E_2-B_2] + \omega_1\omega_2) \Bigr) 
\\
&&{}
-i\omega \frac{2S}{\mathfrak N} T\sum_{q_1 + q_2 = k} \frac{i\omega_1}{[(i\omega_1)^2 - \epsilon_1^2][(i\omega_2)^2 - \epsilon_2^2]}\nonumber\\
&&\times\Bigl(  Q_2^{xz}(Q_1^{xz} + Q_2^{xz}) (E_2-B_2) \\
&&{}
- Q_{\bf k}^{xz} [ Q_1^{xz}(E_2+B_2) + Q_2^{xz}(B_2-E_2) ] \Bigr),
\end{eqnarray}
\end{subequations}
where $k=({\bf k},\omega)$, $q_{1,2}=({\bf q}_{1,2},\omega_{1,2})$ and we drop the index $\bf q$ in Eq.~(\ref{o3}) to light the notation. Sums $\Sigma+\overline\Sigma$ and $\Sigma+\overline\Sigma+\Pi+\Pi^\dagger$ lead to terms (\ref{o3}a)--(\ref{o3}c) and (\ref{o3}d)--(\ref{o3}f), respectively, whereas terms (\ref{o3}g) and (\ref{o3}h) result from $i\omega(\Sigma-\overline\Sigma)$ (see Eq.~(\ref{o})). 

One obtains the following expressions after summation over imaginary frequencies and analytical continuation on $\omega$ from imaginary axis to the real one:
\begin{subequations}
\label{imsum}
\begin{eqnarray}
&&T\sum_{\omega_1} \frac{1}{[(i\omega_1)^2 - \epsilon_1^2][(i\omega_1 - i\omega)^2 - \epsilon_2^2]}
= \nonumber\\
&&\frac{1 + 2N(\epsilon_1)}{2\epsilon_1 [(\epsilon_1+\epsilon_2)^2 - (\omega+i\delta)^2]} 
+
\frac{1 + 2N(\epsilon_2)}{2\epsilon_2 [(\epsilon_1+\epsilon_2)^2 - (\omega+i\delta)^2]}
+
\frac{2(\epsilon_1-\epsilon_2)(N(\epsilon_2)-N(\epsilon_1))}{[(\epsilon_1+\epsilon_2)^2 - (\omega+i\delta)^2][(\epsilon_1-\epsilon_2)^2 - (\omega+i\delta)^2]}
,\\
&& T\sum_{\omega_1} \frac{(i\omega_1)(i\omega-i\omega_1)}{[(i\omega_1)^2 - \epsilon_1^2][(i\omega_1 - i\omega)^2 - \epsilon_2^2]} 
= \nonumber\\
&&\frac{\epsilon_1 (1 + 2N(\epsilon_1))}{2[(\epsilon_1+\epsilon_2)^2 - (\omega+i\delta)^2]} 
+
\frac{\epsilon_2 (1 + 2N(\epsilon_2))}{2[(\epsilon_1+\epsilon_2)^2 - (\omega+i\delta)^2]}
-
\frac{2\epsilon_1 \epsilon_2(\epsilon_1-\epsilon_2)(N(\epsilon_2)-N(\epsilon_1))}{[(\epsilon_1+\epsilon_2)^2 - (\omega+i\delta)^2][(\epsilon_1-\epsilon_2)^2 - (\omega+i\delta)^2]}
,\\
&& T\sum_{\omega_1} \frac{i\omega_1}{[(i\omega_1)^2 - \epsilon_1^2][(i\omega_1 - i\omega)^2 - \epsilon_2^2]} =
\nonumber\\ 
&&\omega\left(\frac{1 + 2N(\epsilon_2)}{2\epsilon_2 [(\epsilon_1+\epsilon_2)^2 - (\omega+i\delta)^2]}
+
\frac{2\epsilon_1 (N(\epsilon_2)-N(\epsilon_1))}{[(\epsilon_1+\epsilon_2)^2 - (\omega+i\delta)^2][(\epsilon_1-\epsilon_2)^2 - (\omega+i\delta)^2]}\right).
\end{eqnarray}
\end{subequations}
We calculate now the real part of $\Omega^{(3)}(\omega,{\bf k})$ using Eqs.~(\ref{imsum}). (\ref{o3}d) is the only term remaining finite at $\omega,{\bf k}=0$ and leading to the contribution (\ref{o3t}) to the spin-wave gap and to the second term in the square brackets in Eq.~(\ref{ginf}). The rest corrections in Eq.~(\ref{o3}) are much smaller than either (\ref{o3}d) or corrections from the Hartree-Fock diagram given by Eqs.~(\ref{o4t<}) and (\ref{o4t>}). Let us estimate them. Term (\ref{o3}a) is of the order of 
$\omega_0^{3/2} k^{5/2}\sin^22\phi_{\bf k} (k\sqrt{\omega_0D} + T)/\sqrt D$, 
$\omega_0^2 k^2\sin^22\phi_{\bf k} (k\sqrt{S\omega_0/D} + T/D)$ 
and 
$\omega_0^2\sin^22\phi_{\bf k} (\omega_0^2 + k^2TD)/D^2$ 
at 
$k\ll S\omega_0/D$, 
$S\omega_0/D\ll k\ll\sqrt{S\omega_0/D}$ 
and 
$k\gg\sqrt{S\omega_0/D}$, 
respectively. Term (\ref{o3}c) is of the order of $k^2\omega_0^2$. It is much smaller than the first term in Eq.~(\ref{o4t<}) if $\ln(D/\omega_0)\gg1$ that we assume to be held. Contributions (\ref{o3}b) cannot give more than (\ref{o3}a) and (\ref{o3}c). Term (\ref{o3}e) is of the order of 
$\omega_0^{5/2}\sqrt k \sin^22\phi_{\bf k} (k\sqrt{\omega_0D} + T)/D^{3/2}$,
$\omega_0^3 \sin^22\phi_{\bf k} (k\sqrt{\omega_0D} + T)/D^2$ 
and 
$\omega_0^3 \sin^22\phi_{\bf k} (\omega_0^2/k^2 + TD)/D^3$ 
at 
$k\ll S\omega_0/D$, 
$S\omega_0/D\ll k\ll\sqrt{S\omega_0/D}$ 
and 
$k\gg\sqrt{S\omega_0/D}$, 
respectively. (\ref{o3}f) does not exceed (\ref{o3}e). (\ref{o3}g) is of the order of $\omega^2(\omega_0^2/D^2) (\ln(D/\omega_0)+T/(\omega_0 + Dk^2)) $. (\ref{o3}h) does not exceed the sum of (\ref{o3}b) and (\ref{o3}f).

Let us turn now to the imaginary part of $\Omega^{(3)}(\omega,{\bf k})$. Corresponding calculations have been done straightforwardly using Eqs.~(\ref{o3}) and (\ref{imsum}) with the following results. 

$k\gg\sqrt{S\omega_0/D}$. Terms (\ref{o3}c) and (\ref{o3}g) give the main equal contributions at small temperature  leading to $T$-independent terms in Eq.~(\ref{damq>}). Terms (\ref{o3}c) and (\ref{o3}g) give equal contributions to the last term in Eq.~(\ref{damq>}). The second term in the square brackets in Eq.~(\ref{damq>}) containing $f(\phi_{\bf k})$ originates from (\ref{o3}a), (\ref{o3}b) and (\ref{o3}c): endowments of (\ref{o3}a) and (\ref{o3}c) are equal and twice as little as that of (\ref{o3}b). The origin of the function $f(\phi_{\bf k})$ is the following. Sum of the form ${\mathfrak N}^{-1}\sum_{\bf q}\delta(\cos(\phi_{\bf k}-\phi_{\bf q}) - \epsilon_{\bf q}/(2Dkq))/(q\epsilon_{\bf q}^2)$ appears in the corresponding expressions in which summation over small momenta $q\ll S\omega_0/D$ is essential. We have for the square of the spectrum at such $\bf q$ (cf.\ Eq.~(\ref{f}))
\begin{equation}
\epsilon_{\bf q}^2 \approx \frac{\alpha(S\omega_0)^3}{4D}
\left(
\tilde q^2 + \tilde q \sin^2\phi_{\bf q} + \frac{4D\Delta^2}{\alpha S^3\omega_0^3}
\right),
\end{equation}
where $\tilde q = q 2D/(S\omega_0)$.

$S\omega_0/D\ll k\ll\sqrt{S\omega_0/D}$. One finds Eq.~(\ref{damq<}) in this regime. Term in Eq.~(\ref{damq<}) containing ${\cal A}_{1,2,3}({\bf q})$ comes from (\ref{o3}d)--(\ref{o3}f) whereas those containing ${\cal B}_1({\bf k})$, ${\cal B}_2({\bf k})$ and ${\cal B}_3({\bf k})$ originate from (\ref{o3}c), (\ref{o3}g) and (\ref{o3}d), respectively. Term in Eq.~(\ref{damq<}) proportional to ${\cal B}_4$ comes from (\ref{o3}d)--(\ref{o3}f).

We demonstrate now in what way the quantity ${\cal L}(q,{\bf k})$ appears in Eq.~(\ref{damq>}). When $S\omega_0/D\ll k\ll\sqrt{S\omega_0/D}$ it is not sufficient to use the leading term in the expression for the spectrum assuming that $\epsilon_{\bf k} \approx \sqrt{S\alpha\omega_0D}k$. Really, we have in this case $\delta(\epsilon_{\bf k} - \epsilon_{\bf q} - \epsilon_{\bf k+q})\propto \delta((\phi_{\bf k} - \phi_{\bf q} - \pi)^2)$ and $\delta(\epsilon_{\bf k} - \epsilon_{\bf q} + \epsilon_{\bf k+q})\propto \delta((\phi_{\bf k} - \phi_{\bf q})^2)$ if $S\omega_0/D\ll k,q,|{\bf k+q}|\ll\sqrt{S\omega_0/D}$. At the same time expressions under the sums do not vanish at $\phi_{\bf q} = \phi_{\bf k}$ or $\phi_{\bf q} = \phi_{\bf k}\pm\pi$. Then, the appearance of the squares in arguments of delta-functions signifies that one should take into account smaller terms in the expression for the spectrum:
\begin{equation}
\label{specm}
\epsilon_{\bf q} \approx \sqrt{S\alpha\omega_0D} 
\left(
q + \frac{S\omega_0}{4D} \sin^2\phi_{\bf q} + \frac{Dq^3}{2\alpha S\omega_0} + \frac{\Delta^2}{2\alpha S\omega_0Dq}
\right).
\end{equation}
Using Eq.~(\ref{specm}) we have if $S\omega_0/D\ll k,q,|{\bf k+q}|\ll\sqrt{S\omega_0/D}$
\begin{equation}
\label{dep}
\epsilon_{\bf k} - \epsilon_{\bf q} \pm \epsilon_{\bf k+q} \approx -\frac{\sqrt{S\alpha\omega_0D}}{2} \frac{k}{q(k-q)}
\left(
q^2_\perp - \frac{3Dk^4}{16\alpha S\omega_0}{\cal L}\left(2\frac qk -1,{\bf k}\right)
\right),
\end{equation}
where ${\bf q}_\perp({\bf q}_\|)$ is the component of $\bf q$ perpendicular (parallel) to $\bf k$. We assume in Eq.~(\ref{dep}) that 
$k-|q_\|| \pm |k+q_\|| = 0$ and $q_\perp\ll q_\|,k$.

$k\ll S\omega_0/D$. In this regime ${\rm Im}\Omega^{(3)}(\omega,{\bf k})$ is finite at high temperature only and one has Eq.~(\ref{damq<<}) for it. Term (\ref{o3}d) only contributes to Eq.~(\ref{damq<<}).

\section{Discussion of the discrepancy between equation (\ref{mag}) for the magnetization and the previous result}
\label{apmag}

In this appendix we comment on the discrepancy between Eq.~(\ref{mag}) for the magnetization and the corresponding expression in Ref.~\cite{mal}. Eq.~(\ref{mag}) coincides with that obtained in Ref.~\cite{mal} at $T\gg {S\omega_0}/{\ln (D/[S\omega_0])}$ and at $T=0$. Meantime, a term was obtained in Ref.~\cite{mal} proportional to $T^{3/2}\omega_0^{-7/4}$ at $T\ll\omega_0^{3/2}/\sqrt J$ that is much larger than the $T$-independent term in Eq.~(\ref{mag}) at $\omega_0^{3/2}/\sqrt J\gg T\gg J(\omega_0/J)^{11/6}$. I believe that the term proportional to $T^{3/2}\omega_0^{-7/4}$ observed in Ref.~\cite{mal} is an artifact. Careful calculation with the bare spectrum (\ref{spec1}) gives the term proportional to $T^{3/2}\omega_0^{-3/4}$ rather than $T^{3/2}\omega_0^{-7/4}$ at $T\ll\omega_0^{3/2}/\sqrt J$ which is much smaller than the $T$-independent term. To confirm this finding we note that the values of $(S - \langle S^z\rangle)/S$ obtained in the regimes of $T\gg\omega_0^{3/2}/\sqrt J$ and $T\ll\omega_0^{3/2}/\sqrt J$ should be of the same order at $T\sim\omega_0^{3/2}/\sqrt J$. At the same time $T^{3/2}\omega_0^{-7/4}\sim\sqrt{\omega_0}$ and $T\ln(T/\omega_0^{3/2})\sim\omega_0^{3/2}$ at $T\sim\omega_0^{3/2}/\sqrt J$. Surprisingly, this artifact has not been revealed so far (see, e.g., Refs.~\cite{rev,2dtc}). To conclude, one leads to the same expression (\ref{mag}) for $(S - \langle S^z\rangle)/S$ as a result of calculations using the bare (\ref{spec1}) and the renormalized spectra.

\section{Effect of the exchange anisotropy}
\label{anis}

We discuss briefly in this appendix the effect of the exchange anisotropy given by Eq.~(\ref{ang}). We consider here only the exchange anisotropy which differs from the on-site one having the form $A\sum_l (S_l^y)^2$ that exists in thin ferromagnetic films. \cite{rev} The reason is that it is technically easier to discuss the exchange anisotropy (\ref{ang}). Moreover it is believed that these two types of anisotropies lead to similar physical results (see Ref.~\cite{rev-jen} and references therein). A more detailed discussion of the effect of the anisotropy in 2D FM with dipolar forces will be published elsewhere. We consider below both signs of $A_{lm}$, i.e. both easy-axis and easy-plane anisotropy. Easy-axis one competes with the dipolar anisotropy which favors in-plane spins alignment. We restrict ourself here to the case of not too large easy-axis anisotropy at which spins lie within the plane. 

Adding ${\cal H}_a$ to the Hamiltonian (\ref{ham0}) one obtains after the Dyson-Maleev transformation (\ref{dm}) a renormalization of $E_{\bf k}$ and $B_{\bf k}$ in the bilinear part of the Hamiltonian (\ref{h2}) 
\begin{equation}
\label{ren2}
E_{\bf k}\mapsto E_{\bf k} + \frac{SA_{\bf k}}{2}, 
\quad 
B_{\bf k}\mapsto B_{\bf k} - \frac{SA_{\bf k}}{2}
\end{equation}
and a contribution to the four-magnon term
\begin{equation}
\label{h4a}
{\cal H}_4^{(a)} = \frac{1}{4 \mathfrak N}\sum_{{\bf k}_1 + {\bf k}_2 + {\bf k}_3 + {\bf k}_4 = 0} 
A_2 a^\dagger_{-1} \left(a_2 - a^\dagger_{-2}\right)a_3a_4.
\end{equation}
As a result of renormalization (\ref{ren2}) the bare spectrum has the form at $k\ll1$ (cf.\ Eq.~(\ref{spec1}))
\begin{equation}
\label{spec0a}
\epsilon_{\bf k}^{(a)} = \sqrt{\left(Dk^2 + S\tilde\omega_0\alpha \right)\left(Dk^2 + \frac{S\omega_0}{2}k\sin^2\phi_{\bf k}\right)},
\end{equation}
where $\tilde\omega_0$ is given by Eq.~(\ref{ot}).
Notice that in the case of easy-axis anisotropy ($A<0$) the spectrum $\epsilon_{\bf k}^{(a)}$ becomes imaginary at small enough $k$ if $\tilde\omega_0<0$. At the same time the in-plane spin alignment becomes energetically unfavorable if the anisotropy is as large as $\tilde\omega_0\le0$. We imply below that $\tilde\omega_0\sim\omega_0$ for $A<0$.

One leads to the following results after the corresponding calculations. Two regimes should be considered in this case: $T\ll S\tilde\omega_0$ and $T\gg S\tilde\omega_0$. We have at $T\ll S\tilde\omega_0$ for ${\rm Re}\Omega(\omega,{\bf k})$ expression (\ref{ot1}) which should be multiplied by $\tilde\omega_0/\omega_0$. As a result the spin-wave gap has the form 
\begin{equation}
\label{adelta}
\Delta^{(a)} = \sqrt{\alpha CS^2\frac{\tilde\omega_0\omega_0^2}{D}},
\end{equation}
i.e., one leads to Eq.~(\ref{delta}) multiplied by $\sqrt{\tilde\omega_0/\omega_0}$. At $T\gg S\tilde\omega_0$ we obtain Eq.~(\ref{ot2}) in which the last two terms should be multiplied by $\tilde\omega_0/\omega_0$. In the case of large $S$ and $T$ discussed in Sec.~\ref{ls} we have for the spin-wave gap expression (\ref{ginf}) multiplied by $\sqrt{\tilde\omega_0/\omega_0}$.

\end{document}